\documentclass[aps,preprint,nofootinbib,superscriptaddress,showpacs]{revtex4}

\usepackage{graphicx}  
\usepackage{amsmath}
\usepackage{amsfonts}
\usepackage{bm}  
\usepackage{color}
\usepackage{slashed}
\usepackage{ulem}

\newcommand{\simle}{\hspace*{0.2em}\raisebox{0.5ex}{$<$}
     \hspace{-0.8em}\raisebox{-0.3em}{$\sim$}\hspace*{0.2em}}

\newcommand{\bea}{\begin{eqnarray}}
\newcommand{\eea}{\end{eqnarray}}

\newcommand{\be}{\begin{equation}}
\newcommand{\ee}{\end{equation}}
\newcommand{\bqa}{\begin{eqnarray}}
\newcommand{\eqa}{\end{eqnarray}}

\newcommand{\eq}[1]{Eq.~\eqref{eq:#1}}
\newcommand{\eqs}[2]{Eqs.~\eqref{eq:#1} and \eqref{eq:#2}}

\newcommand{\fig}[1]{Fig.~\ref{fig:#1}}
\newcommand{\figs}[2]{Figs.~\ref{fig:#1} and \ref{fig:#2}}

\newcommand{\nn}{\nonumber}

\def\mqo2{{\!\!\!}}

\newcommand{\cA}{\mathcal{A}}

\newcommand{\LICS}{$^{6}\text{Li}\text{-}^{133}\text{Cs}$}
\newcommand{\LIRB}{$^{6}\text{Li}\text{-}^{87}\text{Rb}$}
\newcommand{\KRB}{$^{40}\text{K}\text{-}^{87}\text{Rb}$}
\newcommand{\KCS}{$^{40}\text{K}\text{-}^{133}\text{Cs}$}

\begin{document}

\preprint{LA-UR-17-26065}
\title{The Efimov effect for heteronuclear three-body systems at
  positive scattering length and finite temperature}

\author{Samuel B. Emmons}\email{semmons@vols.utk.edu}
\affiliation{Department of Physics and Astronomy, University of
  Tennessee, Knoxville, Tennessee 37996, USA}

\author{Daekyoung Kang}\email{kang1@lanl.gov}
\affiliation{Theoretical Division, MS B283,
Los Alamos National Laboratory, Los Alamos, New Mexico 87545, USA }
\author{Bijaya Acharya}\email{bacharya@utk.edu}
\affiliation{Department of Physics and Astronomy, University of
  Tennessee, Knoxville, Tennessee 37996, USA}
\author{Lucas Platter}\email{lplatter@utk.edu} 
\affiliation{Department of Physics and Astronomy, University of
  Tennessee, Knoxville, Tennessee 37996, USA}

\affiliation{Physics Division, Oak Ridge National Laboratory, Oak
  Ridge, Tennessee 37831, USA}

\date{\today}

\begin{abstract}
  We study the recombination process of three atoms scattering into an
  atom and diatomic molecule in heteronuclear mixtures of ultracold
  atomic gases with large and positive interspecies scattering
  length at finite temperature. We calculate the temperature
  dependence of the three-body recombination rates by extracting
  universal scaling functions that parametrize the energy dependence
  of the scattering matrix.  We compare our results to experimental
  data for the $^{40}$K-$^{87}$Rb mixture and make a prediction for
  $^{6}$Li-$^{87}$Rb. We find that contributions from higher partial
  wave channels significantly impact the total rate and, in
  systems with particularly large mass imbalance, can even obliterate
  the recombination minima associated with the Efimov effect.
  
\end{abstract}


\smallskip
\maketitle
\newpage

\section{Introduction}
\label{sec:introduction}

At sufficiently low energies, the properties of an ultracold gas of
atoms are determined by the $S$-wave scattering length of the atoms. The
scattering length $a$ is usually of the size of the range of the
interaction $\ell$.  However, there exist systems in nature and in the
laboratory, such as nucleons, halo nuclei, or atoms in an external
magnetic field tuned near a Feshbach resonance, in which
$|a| \gg \ell$ \cite{Hammer2010a}. In this case low-energy two-body
observables can be expressed in terms of the $a$ and the associated
momentum scale $k\sim 1/a$ up to corrections proportional to $k\ell$
and $\ell/a$. Three-body systems of identical bosons with large $a$
exhibit a discrete scaling symmetry characterized by a log-periodic
dependence of observables on an additional parameter
$\kappa_\ast$. This is commonly referred to as the Efimov effect
\cite{Efimov70}.  Perhaps the most striking
manifestation of this effect is the emergence of an infinite sequence
of bound states in the unitary limit where $a\rightarrow\pm\infty$, with
energies\footnote {Throughout this work, we adopt a system of units
  where $\hbar=1$.}
\begin{equation}
\label{eq:efimov_energies}
 E^{(n)}=-\lambda^{2(n_\ast-n)}\,  \frac{\kappa_\ast^{2}}{m}, \qquad n = n_\ast, 
 n_\ast \pm 1, n_\ast \pm 2, \ldots.
\end{equation}
Here $m$ can be any quantity with the dimension of mass and
$\kappa_\ast$ is the binding momentum of the three-body state with
$n=n_\ast$. The scaling factor $\lambda$ depends on the mass ratio of
the particles as well as on whether they are identical or
distinguishable. For identical particles, it is
$\lambda_B\approx 22.694$ \cite{Braaten2006259}.  Numerous experiments
with ultracold atomic gases consisting of identical bosons have
confirmed the existence of the Efimov effect by measuring rates of
loss of trapped atoms due to various three-body recombination
processes~\cite{Kraemer:2006,Gross:2009,pollack:2009,Gross2010}.  The
effect was also confirmed in three distinguishable states of $^6$Li
atoms
~\cite{Ottenstein:2008,Huckans:2008fq,Williams:2009,Schmidt:2008fz,NU:2009,
  Braaten:2008wd,Braaten2010,Hammer2010}. Overall, these experiments
have found excellent agreement with theoretical calculations on many
of the important qualitative and quantitative details of Efimov
physics~\cite{0034-4885-80-5-056001}.

There has been a recent
trend~\cite{Barontini:2009,PhysRevLett.113.240402,PhysRevLett.112.250404}
towards performing experiments with heteronuclear systems consisting
of two species of atoms with a large interspecies scattering length,
where $\lambda$ can be driven away from $\lambda_{B}$
\cite{Braaten2006259}. Using light-heavy mixtures thus engenders a
more precise and detailed understanding of Efimov physics by making a
larger number of Efimov states experimentally accessible. Theoretical
studies of Efimov physics in such systems have been performed with
zero-range interactions for zero~\cite{Helfrich2010} and
large~\cite{Zinner2015} intraspecies scattering length.  Finite-range
potential models were used in
Refs.~\cite{Wang2012,PhysRevLett.113.213201}.  By extending the
effective-field-theory analysis of Ref.~\cite{Helfrich2010},
model-independent inclusion of the leading corrections due to finite
interaction ranges and intraspecies scattering length was performed in
Ref.~\cite{PhysRevA.94.032702}.  All of the above-mentioned
theoretical studies have focused on the idealized scenario in which
the temperature of the heteronuclear mixture is exactly zero. However, in real
experimental situations the temperature of the gas, though
small, typically ranges from nK to $\mu$K. This introduces an
additional length scale, the thermal de Broglie wavelengths in the
gas, that leads to additional modifications of the discrete scaling
laws. The finite-temperature effects can be taken into account by
generalizing the $S$-matrix formalism developed to calculate
loss-rates for three-boson systems in
Refs.~\cite{Braaten2006259,Braaten2008} to the heteronuclear
system. In Ref.~\cite{Petrov2015}, this was done for systems that do
not support weakly bound two-body subsystems, i.e. when the
interspecies scattering length is negative.  The purpose of the
present work is to study the temperature dependence of three-body
recombination in two-species mixtures of ultracold atomic gases when
the interspecies scattering length is large and positive, leading to
the existence of a shallow diatomic molecule, while the scattering
length between atoms of the same species remains negligible.  We
perform a detailed analysis of the contribution of different partial
waves to the thermal-averaged recombination rate. We present our
results for two systems of experimental interest, $^{40}$K-$^{87}$Rb
and $^{6}$Li-$^{87}$Rb.

The rest of this paper is organized as follows.  In
Sec.~\ref{sec:first-section}, we briefly review the calculation of the
phase shifts for the scattering of an atom by a diatomic molecule
using the Skorniakov--Ter-Martirosian
(STM) integral equation \cite{Skorniakov:1957aa}, which was originally
applied to the scattering of low-energy neutrons by deuterons and has
been widely used in atomic physics to study the low-energy scattering
of atoms by dimers~\cite{Bedaque:1998km,Ji:2011qg}.
Section~\ref{sec:second-section} then details how the formalism of
Refs.~\cite{Braaten2006259,Braaten2008} can be extended to the
heteronuclear case in order to relate the scattering phase shifts to the
universal scaling functions that parameterize the three-body
recombination rates.  Next, we calculate the temperature-dependent
three-body recombination rate constant as a function of the scattering
length and compare to experimental data in
Sec.~\ref{sec:third-section}.  We summarize and present our concluding remarks
in Sec.~\ref{sec:conclusion}.


\section{STM Equation, Scattering Amplitude, and Phase Shifts}
\label{sec:first-section} 

We consider three-body heteronuclear systems ($A_1A_2A_2$) wherein
the interspecies $S$-wave scattering length $a$ is large and positive,
but the scattering length between any identical atoms is negligible. A
diatomic molecule (labeled $D$) formed by the two atoms $A_1$ and
$A_2$ of masses $m_1$ and $m_2$, respectively, with $m_1 < m_2$, then
has a weakly bound state of binding energy $E_D=1/2\mu a^2$, where
$\mu=m_{1}m_{2}/(m_{1}+m_{2})$.  The elastic scattering phase shift
$\delta^{(J)}_{A_2D}(k_{E})$ for the scattering of atom $A_2$ by the
diatomic molecule $D$ at angular momentum $J$ and
$k_{E}=\sqrt{2\mu_{A_2D}(E+E_{D})}$, where $E$ is the three-body
energy, is given by
\begin{align}
\label{eq:phase}
\cA_{J}(k_{E},k_{E};E,\Lambda)=\frac{2\pi}{\mu_{A_2D}}\frac{1}{k_{E}\cot
\delta^{(J)}_{A_2D}(k_{E})-ik_{E}}\,.
\end{align} 
Here, $\mu_{A_2D}=m_{2}(m_{1}+m_{2})/(2m_{2}+m_{1})$ is the reduced
mass of the $A_2D$ system, and the on-shell scattering amplitude
$\cA_{J}(k_{E},k_{E};E,\Lambda)$ can be obtained by solving the
modified STM equation~\cite{Helfrich2010,Helfrich:2011ut}
\begin{align}
\label{eq:STM}
\cA_{J}(p,k;E,\Lambda)=&\frac{2\pi m_{1}}{a\mu^{2}}(-1)^{n}M_{J}(p,k;E)\nn\\
 &+\frac{m_{1}}{\pi\mu}
\int_{0}^{\Lambda}dq\,q^{2}M_{J}(p,q;E)\frac{(-1)^{n}\cA_{J}(q,k;E,\Lambda)}
{-1/a+\sqrt{-2\mu(E-q^{2}/(2\mu_{A_2D}))-i\epsilon}}\,.
\end{align}
 
The kernel function $M_{J}(p,q;E)$, which can be interpreted as the potential generated by the exchange of the light atom in partial wave $J$, 
is given by 
\begin{align}
\label{eq:generalKJ}
M_{J}(p,q;E)=\frac{1}{pq}Q_{J}\left(\frac{p^{2}+q^{2}-2\mu E-i\epsilon}
{2pq\mu/m_{1}}\right),
\end{align}
where $Q_{J}(z)$ are the Legendre functions of the second kind, which can be written in terms of the Legendre polynomials of order $J$ as
\begin{align}
\label{eq:QJ}
Q_{J}(z)=\frac{1}{2}\int_{-1}^{1}dx\frac{P_{J}(x)}{z-x}\,.
\end{align}
The integer $n$ in Eq.~\eqref{eq:STM} is equal to $J$ if the heavy
particle is bosonic and $J+1$ if it is fermionic.  In this work, we
focus on the bosonic case, which is more relevant for current
experiments. For $J\geq1$, the solutions of Eq.~\eqref{eq:STM}, and
consequently the phase shifts obtained from Eq.~\eqref{eq:phase}, are
independent of $\Lambda$ as long as $p,k,1/a \ll \Lambda$ and
$m_2/m_1<38.63$, beyond which the $D$-wave Efimov effect enters
\cite{Efimov73,Kartavtsev2008}.  We restrict ourselves to these limits
in this work.

However, for $J=0$, the scattering amplitude in Eq.~\eqref{eq:STM},
while finite, does not converge as $\Lambda\rightarrow\infty$. In this
scenario, there is a linear relationship between the cutoff $\Lambda$
and the three-body parameter $\kappa_{*}$,\footnote{One may
  alternatively consider 1/$a_{*0}$, the location of a recombination
  minimum, as a three-body parameter.} resulting in a log-periodicity of the amplitude in the
cutoff with a period equal to the system-dependent scaling factor
$\lambda$ \cite{Bedaque:1998kg,Bedaque:1998km,Hammer:2000nf}.
By solving the STM equation for various $\Lambda$ values in the range
$1/a\ll\Lambda_0<\Lambda<\lambda\Lambda_0$ for some $\Lambda_0$, we
obtain a set of phase shifts $\delta^{(0)}_{A_2D}(k_{E})$
corresponding to various values of $\kappa_\ast$.  As we discuss later
in Sec.~\ref{sec:second-section}, the Efimov radial law is then fit to
these phase shifts in order to obtain universal scaling functions that
are cutoff independent.

The kernel of the STM equation has a branch cut in the complex $q$ plane 
for energies above the three-atom threshold.
To circumvent it, we rotate the integration path by an angle
$\phi$ into the fourth quadrant and integrate along a straight line
from the origin to $\Lambda e^{-i\phi}$~\cite{PhysRev.137.B935}.
Unlike in Ref.~\cite{PhysRev.137.B935}, though, it is important to
include the contribution from the arc connecting $\Lambda e^{-i\phi}$
and $\Lambda$ to obtain correct values for the cutoff-dependent
amplitudes.

\section{Recombination Rates and Scaling Functions}
\label{sec:second-section}

\subsection{Rate Constant and Threshold Behavior}
\label{sec:rec}

A system of three atoms ($A_1A_2A_2$), consisting of two atoms of
species 2 with atomic number density $n_{2}$ and one atom of species 1
with number density $n_{1}$, in a shallow trap can leave the trap as
an $A_2D$ pair by undergoing a three-body recombination process.  For
the $A_1A_2A_2$ system, the recombination rate constant $\alpha$ is defined by
\begin{align}
\label{eq:rate}
	\frac{d}{dt}n_{2}=2\frac{d}{dt}n_{1}=-2\alpha\, n_{1}n_{2}^{2}\,.
\end{align}

At $E=0$, the rate constant $\alpha_s$ for recombination into a 
shallow-bound diatomic molecular state with binding energy $E_D$ 
can be numerically evaluated from the $A_2D$ scattering amplitude 
using~\cite{Helfrich2010,PhysRevA.94.032702}
\begin{align}
\label{eq:alpha_s}
\alpha_{s}=4\mu_{A_2D}\sqrt{\frac{\mu_{A_2D}}{\mu}}a^{2}\left|\cA_{0}\left(0,
\frac{1}{a}\sqrt{\frac{\mu_{A_2D}}{\mu}};0\right)\right|^{2}.
\end{align}
Its dependence on the scattering length $a$ is given by the analytic expression~\cite{Helfrich2010}
\begin{eqnarray}
	\label{eq:alpha_thresh}
	\alpha_{s} &=&
	C(\delta)\,\frac{ \sin^{2}\theta_{*0} +\sinh^{2}\eta_{*}}
	{\sinh^{2}(\pi s_{0}+\eta_{*})+\cos^{2}\theta_{*0}}
	\frac{ a^{4}}{m_{1}}\,.
	\end{eqnarray}
The explicit dependence on the mass ratio $\delta=m_1/m_2$ is captured by 
the coefficient 
\begin{equation}
 C(\delta)=64\pi^2 \left[ (1+\delta^2)\phi(\delta)-\sqrt{\delta(2+\delta)} \right]\,,
\end{equation}
where the phase $\phi(\delta)=\arcsin[(1+\delta)^{-1}]$, and the scaling factor $s_0$ is the
solution of the transcendental equation
\begin{equation}
  \label{eq:s0eq}
s_0\cosh[\pi s_0/2]\, \sin[2\phi(\delta) ]-2\sinh[s_0\phi(\delta)]=0
\,.
\end{equation} 
The angle $\theta_{*0}$ is given by
\begin{equation}
 \theta_{*0} = s_{0}\ln(a/a_{*0}),
\end{equation}
where $a_{*0}$ is the value of the scattering length $a$ at a
recombination minimum, and it follows that \eq{alpha_thresh} is a 
log-periodic function of $a$ with the period $\lambda=e^{\pi/s_0}$. The
inelasticity parameter $\eta_\ast$ is introduced by analytically continuing 
the real-valued $\theta_{*0}$ to the complex value $\theta_{*0}+i\eta_*$,
which is formally equivalent to introducing an anti-Hermitian term in the
three-body Hamiltonian \cite{Braaten:2016dsw,Braaten:2016sja}.
This is done to take into
account the modification of $\alpha_s$ by the existence of deeply
bound diatomic molecular states, which are frequently present in
experimental systems.
The value of $a_{*0}$ that corresponds to
a particular cutoff is determined by fitting the expression in
Eq.~\eqref{eq:alpha_thresh} to the numerical results obtained from
Eq.~\eqref{eq:alpha_s} for $\eta_\ast=0$ over a range of $a$ values.
This gives us the proportional relationship between $\Lambda$ and
$1/a_{*0}$ \cite{Braaten2006259} needed for the extraction of the
universal scaling functions.

Additionally, there is a direct contribution to the total three-body  
recombination rate constant $\alpha$ due to the formation of deeply bound
diatomic molecules in the final state.
The threshold expression for this contribution is given by~\cite{Helfrich2010}
\begin{align}
\label{eq:deepthresh}
\alpha_{d}=C(\delta)\,\frac{\coth(\pi s_{0})\cosh(\eta_{*})\sinh(\eta_{*})}
{\sinh^{2}(\pi s_{0}+\eta_{*})+\cos^{2}\theta_{*0}}
\frac{ a^{4}}{m_{1}}\,.
\end{align}
The maximum threshold value of the recombination rate constant
$\alpha_{th}^{max}$ is the sum of the maxima of both the shallow and deep 
molecule rate constants, which occur at $\theta_{*0}=\pi/2$, and is 
\begin{eqnarray}
	\label{eq:alphaMax}
	\alpha_{th}^{max}=C(\delta)\frac{1+\sinh^{2}\eta_{*}
	+\coth(\pi s_{0})\cosh(\eta_{*})\sinh(\eta_{*})}
	{\sinh^{2}(\pi s_{0}+\eta_{*})}\frac{ a^{4}}{m_{1}}\, .
\end{eqnarray}
Equations~\eqref{eq:alpha_thresh}, \eqref{eq:deepthresh}, and \eqref{eq:alphaMax} provide a useful check for our
three-body recombination rate at non zero energy $K_3^{(J)}(E)$, defined
below. 

\subsection{Three-body Recombination and Universal Scaling Functions}
\label{sec:scaling}
The three-body recombination rate at energy $E$, $K_3^{(J)}(E)$, is
related to the $S$ matrix for the inelastic $A_1A_2A_2\to A_2D$
scattering process.  However, through the unitarity of the  
total $S$ matrix that includes both elastic and inelastic contributions, 
we can write 
the recombination rate purely in terms of the $S$ matrix for elastic $A_2D$
scattering,
$S^{(J)}_{A_2D,A_2D} (E) = \exp[2 i
\delta_{A_2D}^{(J)}(E)]$~\cite{Braaten2008}, as
\begin{equation}
  \label{eq:K3-SADAD}
K_3^{(J)}(E) = \frac{128\pi^{2}\mu^{3/2}}{\mu_{A_2D}^{3/2}}\frac{(2J+1)}{x^{4}}
\left(1-|S^{(J)}_{A_2D,A_2D}(E)|^2\right)\frac{a^{4}}{2\mu}~,
\end{equation}
where the dimensionless scaling variable $x$ is $\sqrt{E/E_{D}}$.
This relation is valid in the absence of deeply bound molecules, the
effects of which we take into account later in this subsection. The
detailed derivation of \eq{K3-SADAD} is given in Appendix \ref{app:app1}.

\subsubsection{$J\geq1$:}
For each total orbital angular momentum $J\geq1$, there is one
corresponding real-valued scaling function 
\begin{align}
f_{J}(x)=1-e^{-4\text{Im}\delta_{A_2D}^{(J)}(E)}\,,
\end{align}
which allows us to obtain the $J^\text{th}$ partial-wave contribution 
to the three-body recombination rate.
Generally, only the first few $J$ values are expected to be necessary
before additional contributions to the total rate become negligible.  As we
increase the value of $J$, the numerical method used to calculate the
phase shifts with which we find $f_{J}(x)$ loses accuracy at small
values of $x$, and we need to use the approximate form
\begin{align}
    \label{eq:fJ}
	f_{J}(x)\approx a_{J}x^{2\lambda_{J}+4}+b_{J}x^{2\lambda_{J}+6}~,
\end{align}
for small $x$, where 
$\lambda_{J}=J$~\cite{Braaten2008,Esry2001}, and the coefficients
$a_J$ and $b_J$ are obtained by fitting \eq{fJ} to $f_{J}(x)$ data at low 
$x$ values with small numerical uncertainties.
The energy-dependent three-body recombination rate
$K_{3}^{(J\geq1)}(E)$ is then given by
\begin{align}
\label{eq:KJ}
K_{3}^{(J\geq1)}(E)=\frac{128\pi^{2}\mu^{3/2}}{\mu_{A_2D}^{3/2}}\frac{(2J+1)f_{J}(x)}
{x^{4}}\frac{ a^{4}}{2\mu}\,.
\end{align}

\subsubsection{$J=0$:}
In the $J=0$ channel, the elements of the $S$ matrix for elastic $A_2D$
scattering are related to universal functions $s_{ij}$ of the scaling
variable $x$ using Efimov's radial law~\cite{Braaten2006259}
\begin{align}
\label{eq:Sad}
S_{A_2D,A_2D}^{(J=0)}(E)=s_{22}(x)+\frac{s_{21}(x)^{2}e^{2i\theta_{*0}-2\eta_{*}}}
{1-s_{11}(x)e^{2i\theta_{*0}-2\eta_{*}}}\,.
\end{align}

\begin{figure}
\centerline{\includegraphics[width=\columnwidth,angle=0,clip=true]
{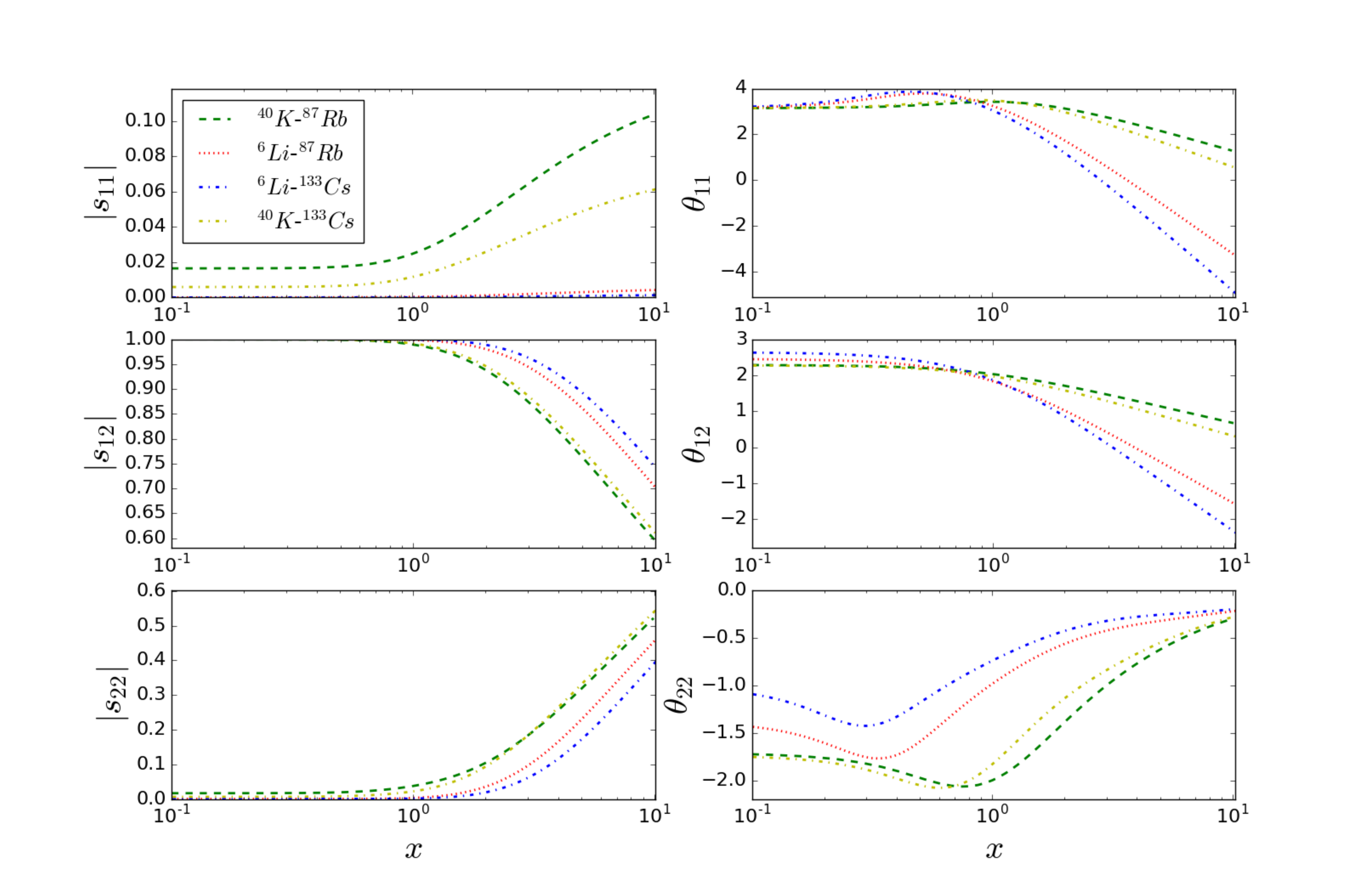}}
\caption{The $S$-wave universal functions for \KRB, \KCS, \LIRB, and \LICS.}
\label{fig:sijKRb}
\end{figure}

We obtain the complex-valued scaling functions $s_{ij}(x)$ by
temporarily setting $\eta_{*}=0$ and fitting \eq{Sad} for each $x$ to
numerical values of phase shifts obtained from Eqs.~\eqref{eq:phase}
and \eqref{eq:STM} for the range of $a_{*0}$ generated by varying
$\Lambda$ as discussed in Sec.~\ref{sec:first-section}.  The $S$-wave
scaling functions of the form $|s_{ij}|e^{i\theta_{ij}}$
for $^{40}\text{K}\text{-}^{87}\text{Rb}$, \LIRB,
\KCS, and \LICS\ are shown in \fig{sijKRb}.
Values for $\eta_{*}$ have been determined or estimated in either
experiments or theoretical calculations for these
systems~\cite{Helfrich2010,Bloom2013,Petrov2015,Ulmanis2016} and are
included in the S-wave three-body recombination rate for shallow and
deep diatomic molecules.

%
With the universal functions $s_{ij}$, we can
calculate the $S$-wave heteronuclear three-body recombination rate
$K_{3}^{(0)}(E)$ from
\begin{align}
\label{eq:Kshallow}
K_{3}^{(0)}(E)=&\frac{128\pi^{2}\mu^{3/2}}{\mu_{A_2D}^{3/2}}\frac{1}{x^{4}}
\Bigg(1-\left|s_{22}(x)+\frac{s_{12}(x)^{2}e^{2i\theta_{*0}-2\eta_{*}}}
{1-s_{11}(x)e^{2i\theta_{*0}-2\eta_{*}}}\right|^{2} \nn\\
&-\frac{(1-e^{-4\eta{*}})
\left|s_{12}(x)\right|^{2}}
{\left|1-s_{11}(x)e^{2i\theta_{*0}-2\eta_{*}}\right|^{2}}\Bigg)
\frac{ a^{4}}{2\mu}\, ,
\end{align}
where the third term in large parentheses in \eq{Kshallow} arises from
incorporating possible transitions from an $A_2D$ scattering state or
three-atom scattering state into an atom and a deeply bound diatomic
molecule in the intermediate state. To obtain results for a given system, we take the position
of one of the recombination minima as an experimental or theoretical
input for that system.

There is an additional contribution from the formation of deeply bound
molecules in the final state, whose significance for a particular system depends on the size
of $\eta_{*}$. These effects are subleading in the zero-range limit
for $J\geq 1$~\cite{Braaten2008}. However, for $J=0$, the
contribution,
\begin{align}
\label{eq:Kdeep}
K_{3}^{deep}(E)=\frac{128\pi^{2}\mu^{3/2}(1-e^{-4\eta_{*}})\left[
1-\left|s_{11}(x)\right|^{2}-\left|s_{12}(x)\right|^{2}\right]}
{\mu_{A_2D}^{3/2}\,x^{4}\left|1-s_{11}(x)e^{2i\theta_{*0}-2\eta_{*}}\right|^{2}}
\frac{ a^{4}}{2\mu}\,,
\end{align}
appears at leading order and must be added to the rate of
recombination into shallow diatomic molecules in order to obtain the
full recombination rate.

We have checked and verified the $E\rightarrow0$ limits of
$K_{3}^{(0)}(E)$ and $K_{3}^{deep}(E)$ given by
Eqs.~\eqref{eq:Kshallow} and \eqref{eq:Kdeep} by comparing them to the
corresponding threshold expressions given by
Eqs.~\eqref{eq:alpha_thresh} and \eqref{eq:deepthresh} multiplied by a
factor of 2 that comes from the statistics of the system.
The total threshold $S$-wave recombination rate containing the
contribution of both shallow and deep states
$K_{3}^{(0)}(0)+K_{3}^{deep}(0)$ has a maximum value of
$K_{th}^{max}$ at $\theta_{*0}=\pi/2$. This is related to
$\alpha_{th}^{max}$ defined in Eq.~\eqref{eq:alphaMax} by the relation
$K_{th}^{max}=2\alpha_{th}^{max}$.

\begin{figure}[t]
\centerline{\includegraphics[width=9cm,angle=0,clip=true]
  {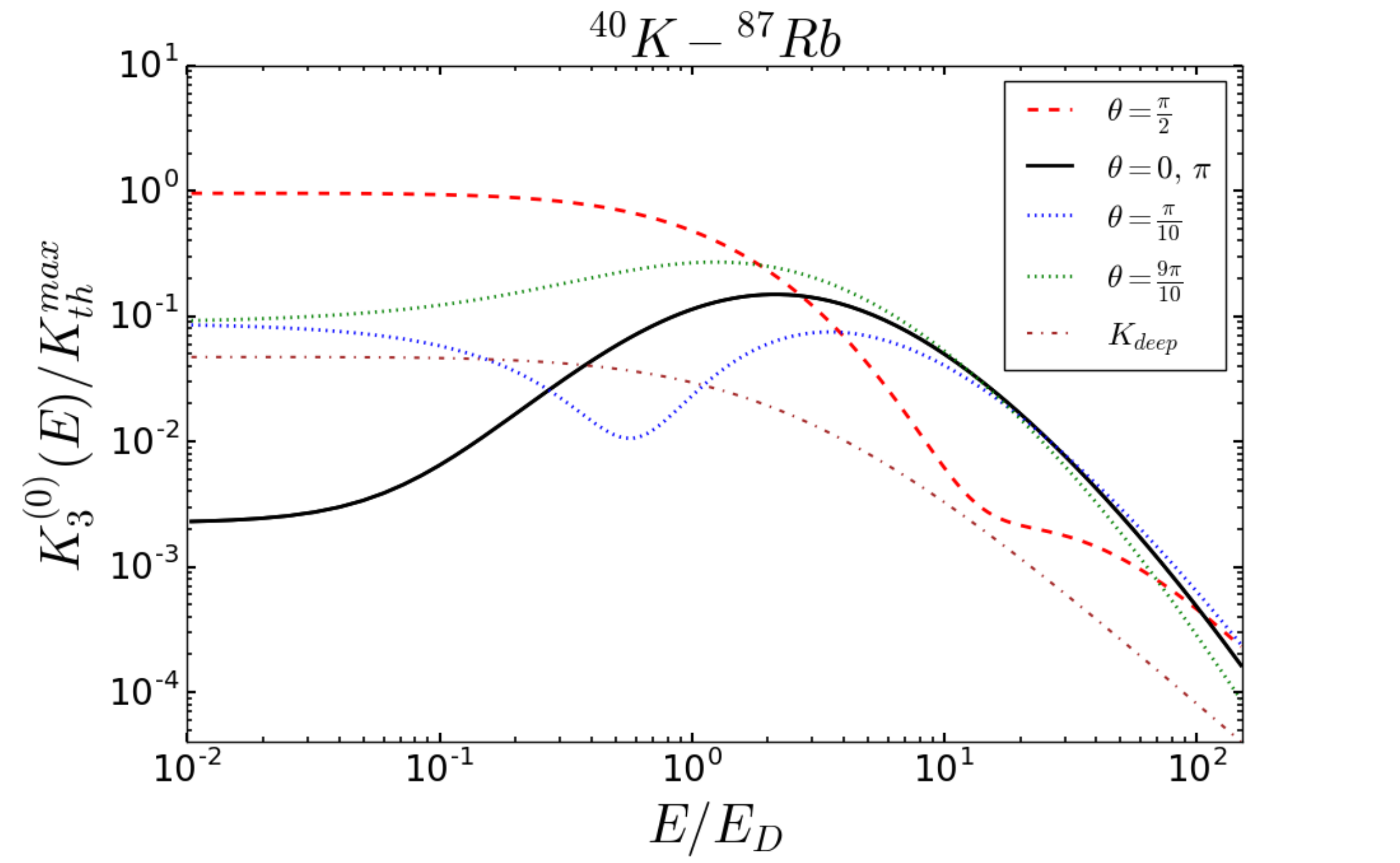}
\includegraphics[width=9cm,angle=0,clip=true]{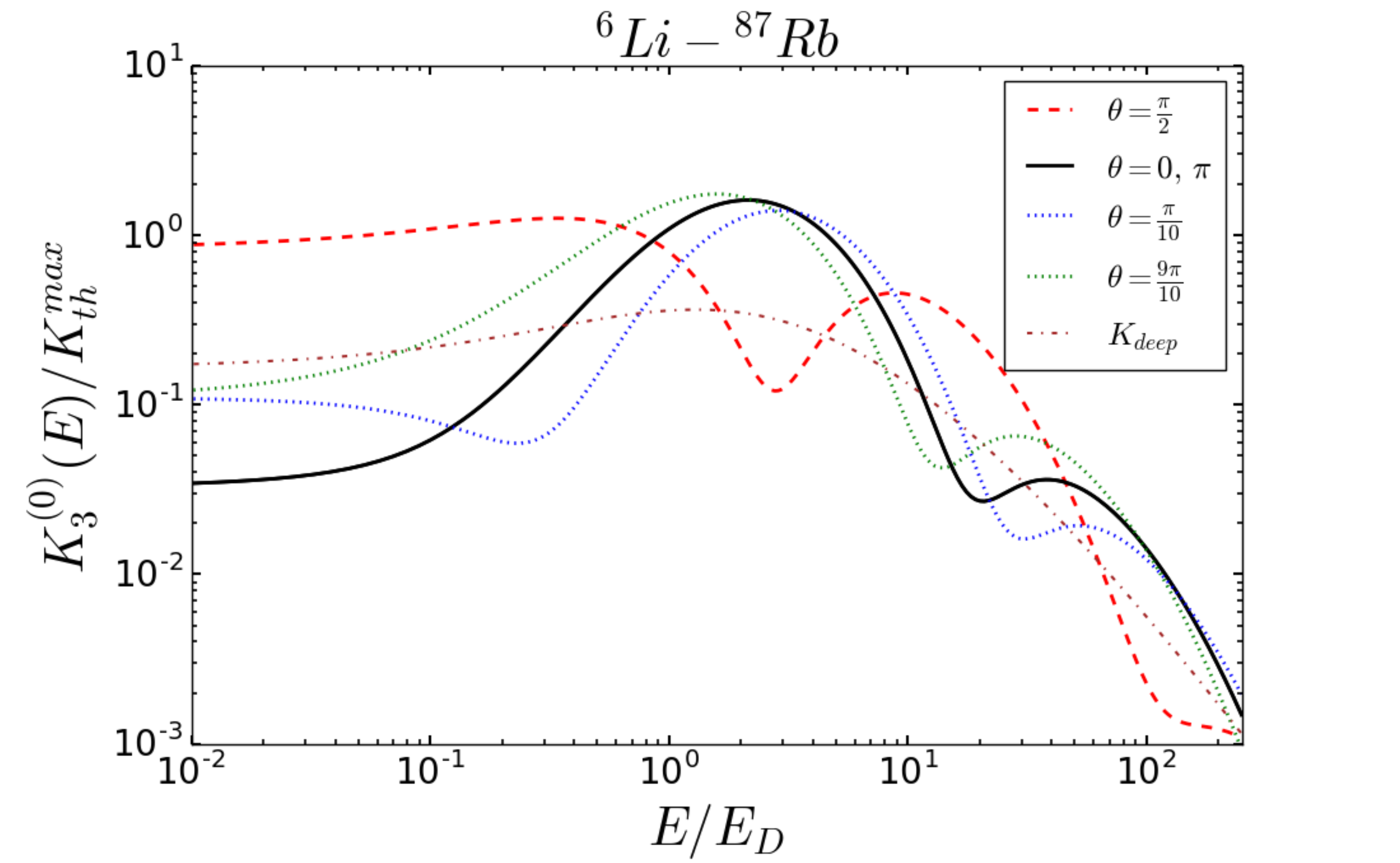}}
\caption{Shown on the left is the $J=0$ recombination rate divided by the
  maximum threshold value $K_{th}^{max}$ for a variety of
  values of $\theta_{*0}$ in the \KRB\ system with $\eta_{*}=0.05$
  and on the right is the $J=0$ recombination rate divided by the
  maximum threshold value, $K_{th}^{max}$, for a variety of
  values of $\theta_{*0}$ in the \LIRB~ system with $\eta_{*}=0.2$. }
\label{fig:K0_KRb}
\end{figure}

In Fig.~\ref{fig:K0_KRb}, we plot the energy dependence of
$K_{3}^{(0)}(E)$ and $K_{3}^{deep}(E)$ at various values of
$\theta_{*0}$ for the \KRB\ and \LIRB\ systems.\footnote{Numerical data 
for these and other systems can be provided by the authors on request.} The rates are
expressed in the units of $K_{th}^{max}$. The variations in the shape
of the $K_{3}^{(0)}(E)$ curves by up to several orders of magnitude
as $\theta_{*0}$ varies show that the energy dependence of
$S$-wave recombination into a shallow diatomic molecular state has an
intricate dependence itself on the scattering length $a$ and the scaling
parameter $s_0$.

\begin{figure} [t]
		\centerline{\includegraphics[width=9cm,angle=0,clip=true]
		{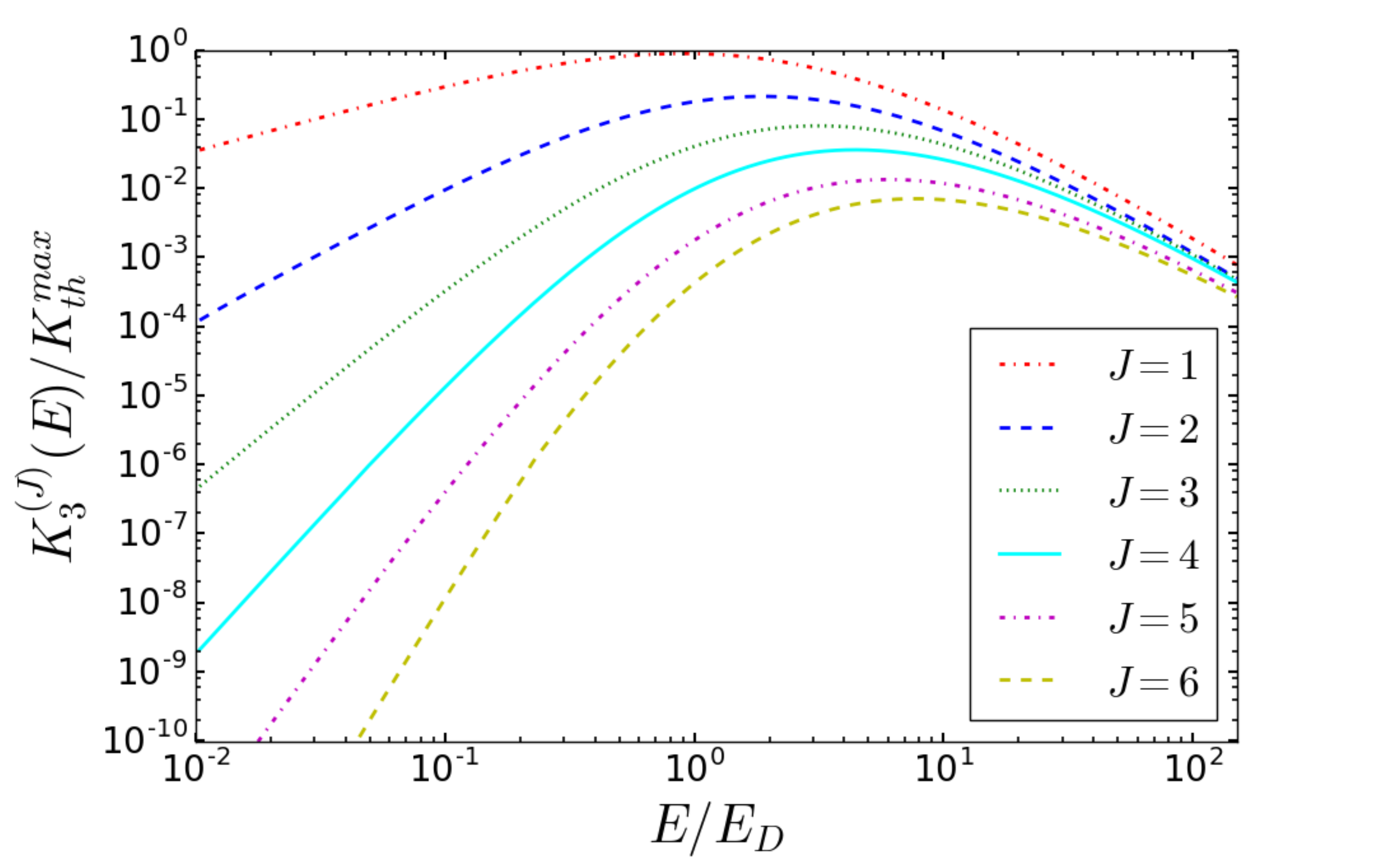}
		\includegraphics[width=9cm,angle=0,clip=true]{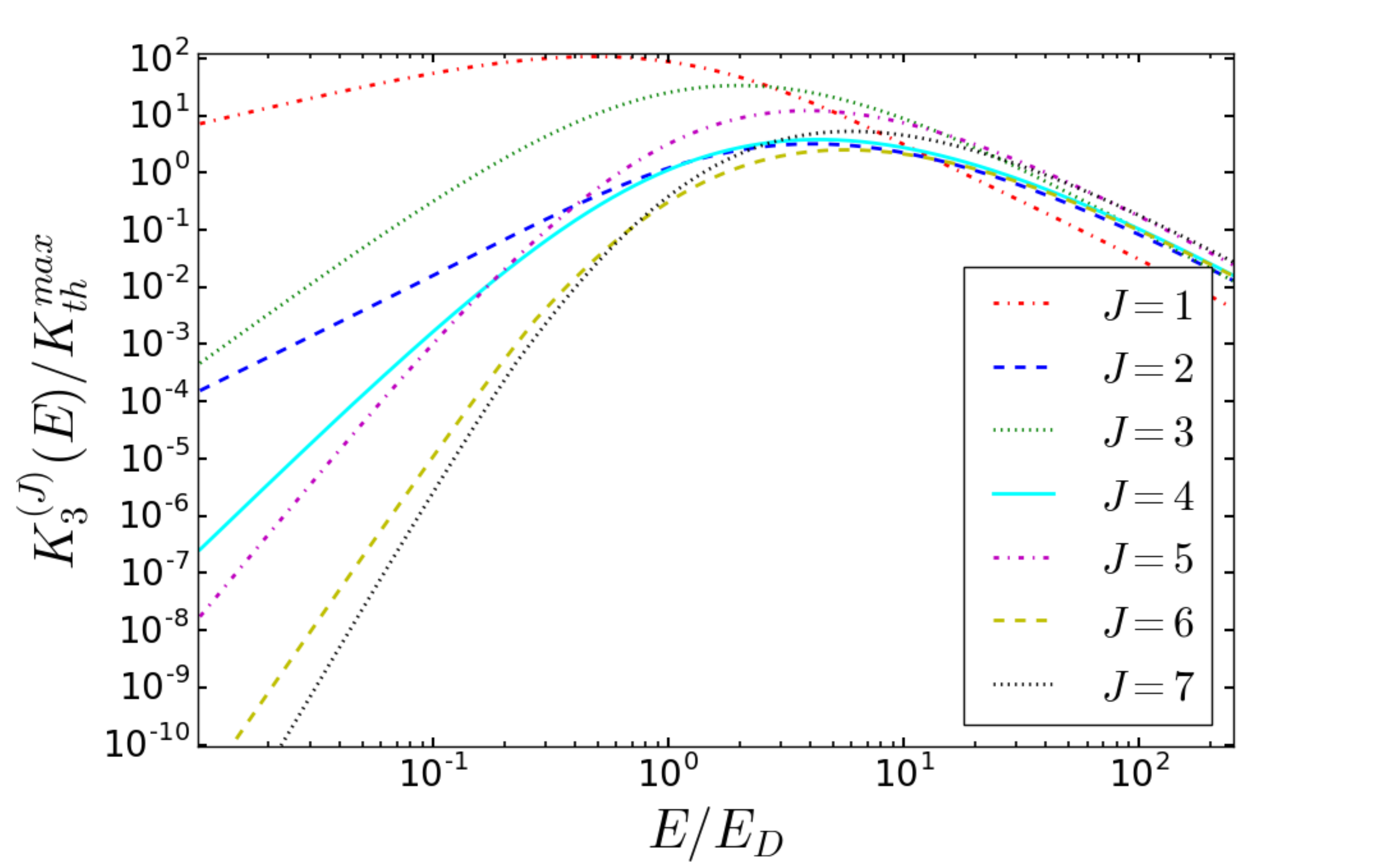}}
	\caption{Shown on the left is $K_{3}^{(J)}(E)/K_{th}^{max}$ for \KRB 
		and on the right is $K_{3}^{(J)}(E)/K_{th}^{max}$ for \LIRB.}
	\label{fig:KJ_KRb}
\end{figure}

Figure~\ref{fig:KJ_KRb} shows the energy dependence of
$K_{3}^{(J\geq1)}(E)$ for the \KRB\ and \LIRB\ systems. These are
expressed in the units of the threshold $S$-wave rate maximum
$K_{th}^{max}$.  For the \KRB\ system, we observe diminishing
contributions as we go to higher partial waves.  This is
different from the behavior of a system of three
identical bosons, in which the contribution of the $J=1$ partial wave
was found to be comparable to that of the $J=4$ partial wave
\cite{Braaten2008}. The near-threshold energy dependence of the
recombination rates $K_{3}^{(J)}(E)$ in Figs.~\ref{fig:K0_KRb} and
\ref{fig:KJ_KRb} agrees with the predictions given in
Ref.~\cite{DIncao2006}. However, we do not reproduce the dependence on
the mass ratio $\delta$ suggested by Ref.~\cite{DIncao2006}.

Comparing \figs{K0_KRb}{KJ_KRb} informs us about the temperature scale
around and above which the recombination minima are unlikely to be
measured due to large partial wave contributions.  In \KRB, the $J=1$
partial wave becomes larger than the $S$ wave around $E_D$ and the
corresponding temperature is
$T_{\rm KRb}= 0.3 E_D/k_B \approx 0.1\, (a/a_0)^{-2}$ K, where
$a_{0}$ is the Bohr radius.
On the other hand, in the \LIRB\ system it happens at a very low energy $\sim 10^{-3} E_D$, 
which corresponds to the 
temperature $T_{\rm LiRb}= 10^{-3} E_D/k_B \approx 0.015\, (a/a_0)^{-2}$ K.
These relations either give a maximum scattering length below which
the minima can be observed, provided that the universal region
$a\gg \ell$ still exists, or set a target temperature below which we
may begin to observe known minima around the value of $a$ and below.

\section{Comparison With Experiment}
\label{sec:third-section}

\begin{figure}
  \centerline{\includegraphics[width=\columnwidth,angle=0,clip=true]
    {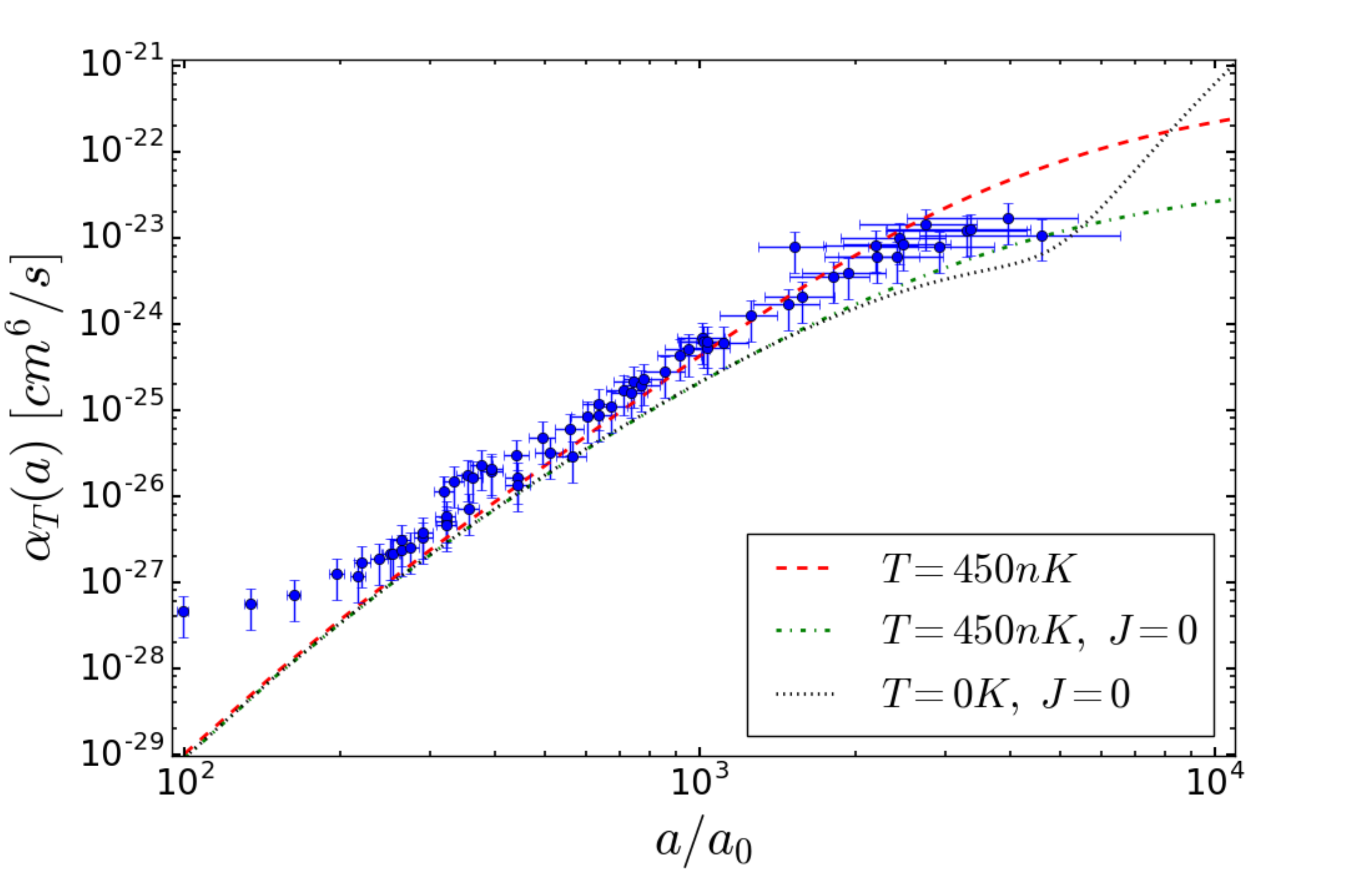}}
  \caption{Recombination rate constant $\alpha_T$ as a function of
    the scattering length $a$ for \KRB\ with
    $\eta_{*}=0.05\pm0.02$~\cite{Helfrich2010} and the three-body
    parameter adjusted to reproduce a recombination minimum at
    $a_{*0}\approx5000a_{0}$. The dashed
     red line at $450$ nK
    corresponds to the average temperature at which the data of Bloom
    {\it et al.} was taken \cite{Bloom2013}.}
	\label{fig:alpha}
\end{figure}

To make a comparison of our results with experiments we require input
values for $a_{*0}$ and $\eta_{*}$.  We calculate the
contributions from all different scattering sectors and combine them
into a total recombination rate,
\begin{equation}
\label{eq:Ktotalshallow}
K_{3}(E)=\sum_{J=0}^{\infty}K_{3}^{(J)}(E)+K_{3}^{deep}(E)\,.
\end{equation}
We then perform a thermal average over $K_{3}(E)$ to obtain the
recombination rate constant for a specific scattering length at a
finite temperature used in relevant experiments~\cite{Braaten2008},
\begin{align}
\alpha_T\approx\frac{\int_{0}^{\infty}dE\,E^{2}e^{-E/(k_{B}T)}K_{3}(E)}
{2\int_{0}^{\infty}dE\,E^{2}e^{-E/(k_{B}T)}}\,,
\end{align}
where the coefficient 2 in the denominator is the symmetry factor.

\begin{figure}
	\centerline{\includegraphics[width=\columnwidth,angle=0,clip=true]
		{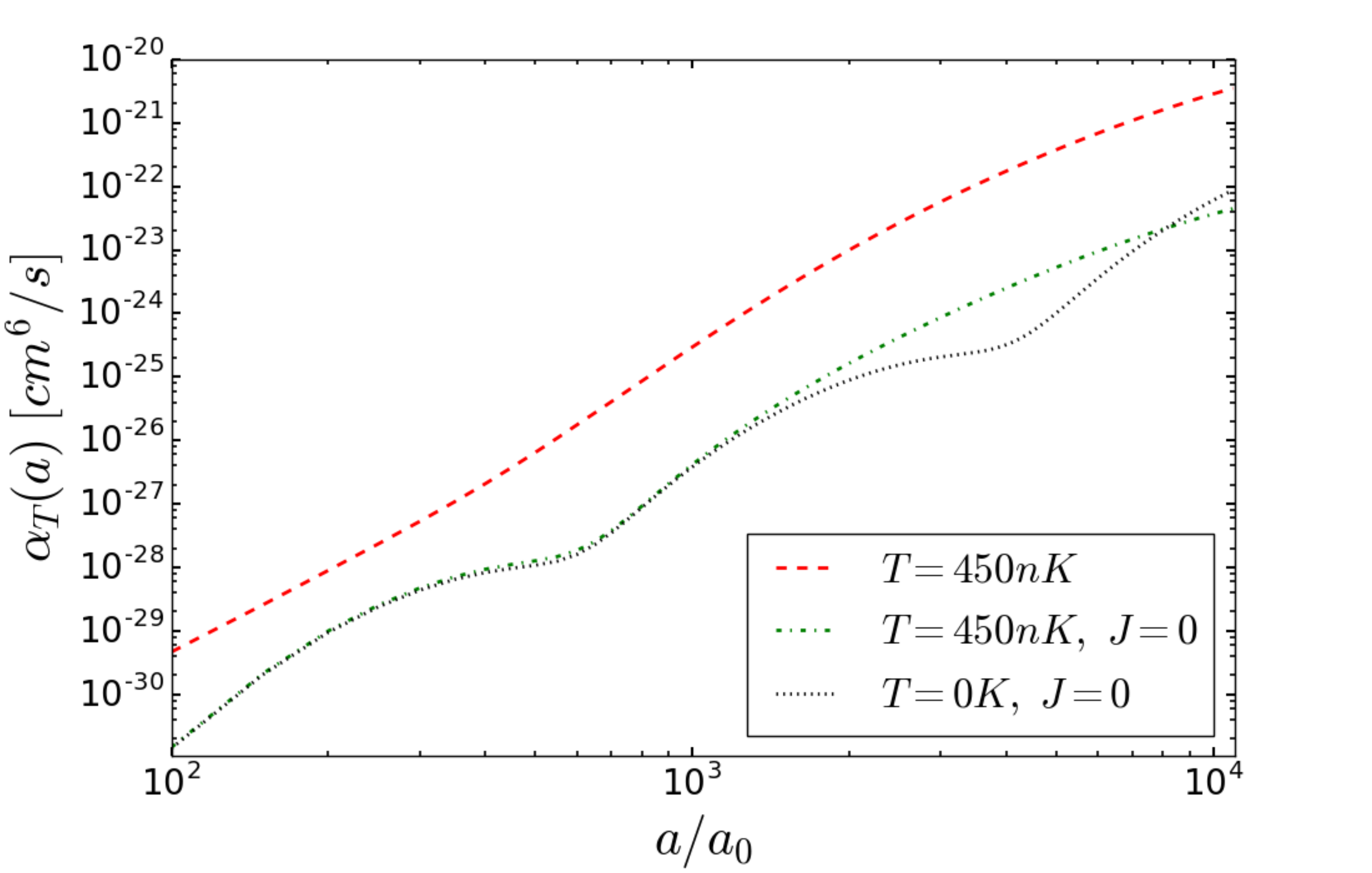}}
              \caption{Recombination rate constant $\alpha_T$ as a
                function of the scattering length $a$ for \LIRB\ for
                $\eta_{*}=0.2$~\cite{Petrov2015} with a recombination
                minimum at $a_{*0}\approx610a_{0}$.}
	\label{fig:alphaLiRb}
\end{figure}
The inelasticity parameter for a \KRB\ mixture was estimated by
Bloom {\it et al.}~\cite{Bloom2013} to be $\eta_{*}=0.26$ by matching a
threshold formula for the atom-molecule relaxation loss rate coefficient $\beta$
to experimental data.  However, they later gave $\eta_{*}=0.02$ as a good match
for their measurements of the rate constant $\alpha$. 
The published data, though, is restricted to $a$ values smaller than the thermal
wavelength of the atoms set by the temperature of the gas in the experiment.
Meanwhile, Helfrich, \textit{et al.} find
$\eta_{*}=0.05\pm0.02$~\cite{Helfrich2010} by fitting their Eq.~(20)
to the corresponding data from Ref. \cite{Zirbel2008}. 
We use the experimental value $a_{*}=230a_{0}\pm30a_{0}$ obtained
in Ref. \cite{Bloom2013} for \KRB~ to determine the position of the
recombination minimum $a_{*0}\approx 5000a_{0}$. We achieve this by
employing the \KRB~universal relation
$a_*/a_{*0} = 0.51 \exp(\pi/2 s_0)$~\cite{Helfrich2010}, which is
exact in the zero-range limit employed in this work.
Here, $a_{*}$ is the value of $a$ where the Efimov trimer state reaches 
the $A_2D$ threshold. On the
other hand, Wang \textit{et al.}~\cite{Wang2012} predicted from a
theoretical calculation that $a_{*0}=2800a_{0}$. In their procedure to
obtain this value, they set the Rb-Rb scattering length to
$a_{22}=100a_{0}$.
The relatively large temperatures used in the experiment by
Bloom {\it et al.} do not allow for the observation of this feature.
Therefore, the discrepancy between the universal prediction obtained
from the value of $a_*$ and the result presented in
Ref.~\cite{Wang2012} cannot be addressed. We find that a temperature
of approximately 10~nK would be necessary to clearly observe recombination minima
in this experiment.

Bloom {\it et al.}  also gave evidence that we can neglect the
${}^{87}$Rb-${}^{87}$Rb-${}^{87}$Rb recombination channel due to the small
scattering length $a_{22}$, with an observed ratio of $^{87}$Rb loss
to $^{40}$K loss of $2.1(1)$ indicating that the dominant loss channel
is $^{40}$K-$^{87}$Rb-$^{87}$Rb recombination. We also note that the
uncertainty introduced 
by neglecting the small scattering length
$a_{22}$ in the calculation of 
$^{40}$K-$^{87}$Rb-$^{87}$Rb recombination rate 
is of the order of $a_{22}/a$. The perturbative approach
introduced in Ref.~\cite{PhysRevA.94.032702} could be employed to
account for such corrections as long as $a_{22}<a$.

Further, though we use the value of $a_{*0}\approx5000a_{0}$ for the
position of a recombination minimum, this
minimum was not probed in Ref.~\cite{Bloom2013}, since they were limited by
their experimental temperature of $T\sim300$ nK to
$a\simle3000a_{0}$.\footnote{Although the average
temperature for their experiment was around $450$ nK, the
data at the largest $a$ values was taken near $300$ nK.}
Experimental uncertainties become quite
large near and beyond this value. This
means that for \KRB\ no Efimov features were definitively observed for
three-body recombination in currently accessed positive scattering
lengths. In \fig{alpha}, we show the data of Ref. \cite{Bloom2013}
and our numerically obtained curves for rate constant $\alpha_T$,
with one curve showing the $J=0$ contribution to the rate constant and
another showing the total rate after summing over $J$. We also include
the zero-temperature result obtained by summing
\eqs{alpha_thresh}{deepthresh} for comparison. In each of our curves
in the figure we have selected $a_{*0}=5000a_{0}$ and $\eta_{*}=0.05$.
The agreement of the $450$-nK curve with the experimental data
is excellent in the large-$a$ region where the neglected contributions
due to finite range and finite $a_{22}$ corrections become
small~\cite{PhysRevA.94.032702}. The size of the discrepancy at
$a\simle200a_0$ suggests that the latter might perhaps be more
important for this experiment than the former, since, with a quoted
value of the van der Waals range $R_{vdW}=72a_{0}$ \cite{Bloom2013},
range corrections are expected to be about 35\%--70\% in this
region. Our results at lower temperatures indicate the minima at 
$5000a_{0}$ can only be observed at temperatures well below 
$10\,{\rm nK}$, which may not be experimentally feasible.

The similar $^{39}\text{K-}^{87}\text{Rb}$ and
$^{41}\text{K-}^{87}\text{Rb}$ systems were studied by Wacker {\it et
  al.}~\cite{Wacker2016}. For $a>0$, no signatures of Efimov
resonances were seen in either mixture for accessible scattering
lengths and temperatures, further demonstrating how a large scaling factor
makes the
observation of universality difficult and giving a compelling argument in
favor of using systems with a larger mass imbalance such as \LIRB\ or
\LICS. We therefore study the effects of temperature on the
recombination rate constant for the \LIRB\ system in \fig{alphaLiRb}.
We examine a couple of different sources to obtain inputs for $a_{*0}$
and $\eta_{*}$.  First, the $^{7}$Li-$^{87}$Rb system was studied by
Maier {\it et al.} in Ref. \cite{Maier2015}, and they found a value of
$\left|a_{-}\right|=1870a_{0}\pm121a_{0}$. They further suggest a value of
$a_{-}$ of $-1600a_{0}$ for \LIRB, which, with $\left|
a_{-}\right|/a_{*0}=\text{exp}(\pi/2s_{0})$, gives a recombination minimum
position of $a_{*0}\approx610a_{0}$. Additionally, for \LIRB, Petrov
and Werner, in the absence of any known experimental results, give
$\eta_{*}=0.2$~\cite{Petrov2015}. We adopt the use of $\eta_{*}=0.2$
and $a_{*0}=610a_{0}$ in \fig{alphaLiRb}. We find that the
recombination minima are obscured by the finite temperature effects,
particularly by the ones that enter in partial waves $J\geq1$, though 
even for the $J=0$ (dash-dotted) line the second minimum is obscured. The
effects of higher partial waves begin to be suppressed below $\sim10$ nK and
the minimum at $610a_{0}$ becomes accessible in experiments.  We have
performed the partial-wave analysis shown earlier in \fig{KJ_KRb} for
several systems and found that higher partial-wave contributions
become increasingly dominant at smaller $m_{1}/m_{2}$. 
Therefore, in order to see the detailed universal behavior, it
appears that one must prepare the system at very low temperatures for
small mass ratios. Illustrative plots of the recombination rate constant for
two additional systems beyond those shown above are given in Appendix
\ref{app:app2}.

This trend stands in contrast to the suggestion of D'Incao and Esry
\cite{DIncao2006}
that the dominant contribution to recombination in systems in which $A_2$ is 
bosonic comes from the $J=0$ channel. While this is certainly true when 
$E=0$, it does not appear to be true at all values of $E$, 
particularly for systems with small $m_1/m_2$.

\section{Conclusion}
\label{sec:conclusion} In this work, we considered three-body
recombination in heteronuclear systems with positive interspecies
scattering length at finite temperature. Using the STM equation, we
obtained sets of universal scaling functions that can be used to
calculate the temperature-dependent recombination rate for arbitrary
values of the three-body parameter and inelasticity parameter
$\eta_*$. Every mass ratio requires a new set of scaling functions and
we calculated these for various systems of interest. We also
calculated the universal scaling functions for higher partial waves
that do not display the Efimov effect but contribute to the total loss
rate. Our results show that observing the Efimov effect becomes
difficult due to relatively large recombination rate contributions
from higher partial-wave scattering channels at experimentally feasible
temperatures. This obfuscation of
$S$-wave universality becomes particularly acute for systems with small
$m_1/m_2$ and reduces their favorability for the experimental
observation of Efimov features when $a>0$.  We have compared our
results with experimental results for three-body combination in an
ultracold mixture of \KRB\ atoms and found good agreement with the
data.

Addressing the impact of corrections due to the
finite range of the interactions is left for future work. 
These effects were studied in the
framework of effective field theory for identical bosons in
Refs.~\cite{Ji:2011qg,Ji:2015hha} and for heteronuclear systems in
Ref.~\cite{PhysRevA.94.032702}. Including range corrections to the
temperature-dependent three-body recombination process will enable us
to understand Efimov physics even when $a$ is not particularly large
and might help us avoid the range of $a$ values where higher partial-wave
contributions are dominant.  The effects of a finite intraspecies
scattering length $a_{22}$ have been incorporated
perturbatively~\cite{PhysRevA.94.032702} for
$\vert a_{22}\vert\ll \vert a\vert$ and $T=0$ K and
nonperturbatively~\cite{Petrov2015} for
$\vert a_{22}\vert\sim \vert a \vert$ at finite $T$ for $a<0$; however,
this remains to be done for finite $T$ when $a>0$. Major extensions to
the existing formalism will be required to accommodate additional
scattering channels if $a_{22}>0$.

\section*{Acknowledgments}
We thank Ruth Bloom for providing us with the recombination data of
Ref. \cite{Bloom2013} and Eric Braaten for comments on the
manuscript. This work was supported by the U.S. Department of Energy
through the Office of Science, Office of Nuclear Physics under
Contracts No. DE-AC52-06NA25396 and No. DE-AC05-00OR22725, an Early Career
Research Award, the LANL/LDRD Program, and the National Science
Foundation under Grant No. PHY-1555030.

\begin{appendix}


\section{Phase Space Factors}
\label{app:app1}

We calculate the three-body recombination rate by relating it to the
cross section for inelastic $A_{2}D$ scattering $\sigma_{A_2D}^{\rm
  (inelastic)}$. This cross section is defined as 
\begin{equation}
  \sigma_{A_2D}^{\rm (inelastic)} = \frac{1}{2 v_{A_2D}} | \mathcal{A}_{A_2D,A_1A_2A_2}|^2\,  \Phi_3~,
\end{equation}
where $\mathcal{A}_{A_2D,A_1A_2A_2}$ denotes the amplitude for a 
transition from an $A_{2}D$ state to three atoms, 
the relative velocity of the atom $A_2$ and molecule $D$ is
$v_{A_2D}=k_E/\mu_{A_2D}$, where $k_E=\sqrt{2\mu_{A_2D}( E+E_D)}$,
and the flux factor $\Phi_3$ is the three-body phasespace.
We also include a symmetry factor of 2 into the expression for
the total cross section since we have two identical particles in the final
state.

Further, one can write the three-body recombination rate $K_3$ as
\begin{equation}
  \label{eq:K3-amplitude}
  K_3 = |\mathcal{A}_{A_1A_2A_2, A_2D}|^2 \Phi_2
  =2 v_{A_2D}\,\frac{\Phi_2}{\Phi_3}\, \sigma_{A_2D}^{\rm (inelastic)}~.
\end{equation}
The inelastic cross section can be rewritten in terms of total and elastic ones as
\begin{eqnarray}
\sigma_{A_2D}^{\rm (inelastic)} 
&=& \sigma_{A_2D}^{\rm (tot)}-\sigma_{A_2D}^{\rm (elastic)}
\nn\\
&=&(2J+1)\left[ \frac{2\mu_{A_2D}}{k_E}\text{Im}\, A_J(k_E,k_E,E) - \frac{\mu_{A_2D}^2}{\pi} |A_J(k_E,k_E,E)|^2\right]
\nn\\
&=&(2J+1)\frac{\pi}{k_E^2}\left[ 1-\left\vert e^{2i\delta^{(J)}_{A_2D}(E)} \right\vert^2 \right]~,
\end{eqnarray}
where we used \eq{phase} to arrive at the last line.
This relates the recombination rate to the phase shift 
(i.e. the $S$ matrix element) given in \eq{K3-SADAD}
with a normalization factor determined by the ratio $\Phi_2/\Phi_3$.

The two-body phasespace $\Phi_2$ is given by 
\begin{eqnarray}
\Phi_2 &=& \int \frac{\hbox{d}^3 p_A}{(2\pi)^3}\frac{\hbox{d}^3 p_D}{(2\pi)^3}
(2\pi)^3\delta^{(3)}(p_A+p_D)\,
2 \pi \delta \left(E - \frac{p_A^2}{2 m_2}-\frac{p_D^2}{2(m_1+m_2)}+E_D\right)
\nn\\
&=&\frac{\mu_{A_2D}\, k_E}{\pi}~.
\end{eqnarray}
The three-atom final state phasespace factor is
\begin{eqnarray}
\Phi_3 &=&\int\prod_{i=1}^{3}\frac{\hbox{d}^3 p_i}{(2\pi)^3} 
(2\pi)^3\delta^{(3)}({\bf p}_1+{\bf p}_2+{\bf p}_3)\, 2\pi \delta\left(E-\frac{p_1^2}{2m_1}-\frac{p_2^2}{2m_2}-\frac{p_3^2}{2m_2}\right) 
\nn\\
&=&\frac{(\mu \mu_{A_2D})^{3/2}}{8\pi^2} E^2~,
\end{eqnarray}
where the ${\bf p}_i$, with $i=1,2,3$, denote the momenta
of the three final state atoms.
Using these phase-space factors in Eq.~\eqref{eq:K3-amplitude} leads to
the final result
\begin{equation}
  \label{eq:K3sigma_final}
  K_3 = \frac{16
    \pi^2}{(\mu\mu_{A_2D})^{3/2} E^2} (2J+1)\left[
    1-|e^{2 i \delta_{A_2D}^J}|^2\right]~.
\end{equation}
Making the substitution $E=x^2 /2\mu a^2$ leads to Eq.~\eqref{eq:K3-SADAD}~.

\section{Additional Systems}
\label{app:app2}
\begin{figure} [h]
		\centerline{\includegraphics[width=9cm,angle=0,clip=true]
		{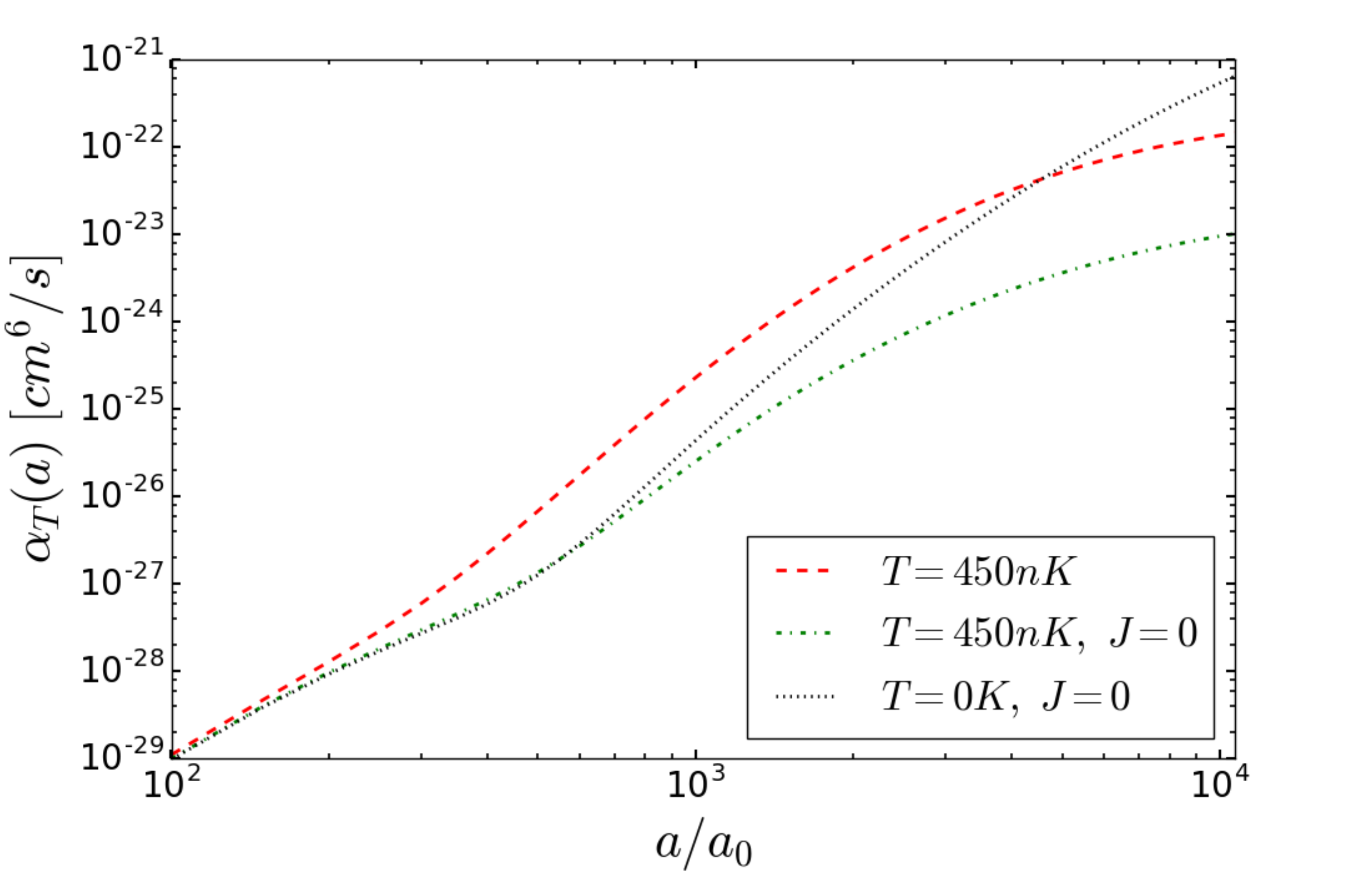}
		\includegraphics[width=9cm,angle=0,clip=true]{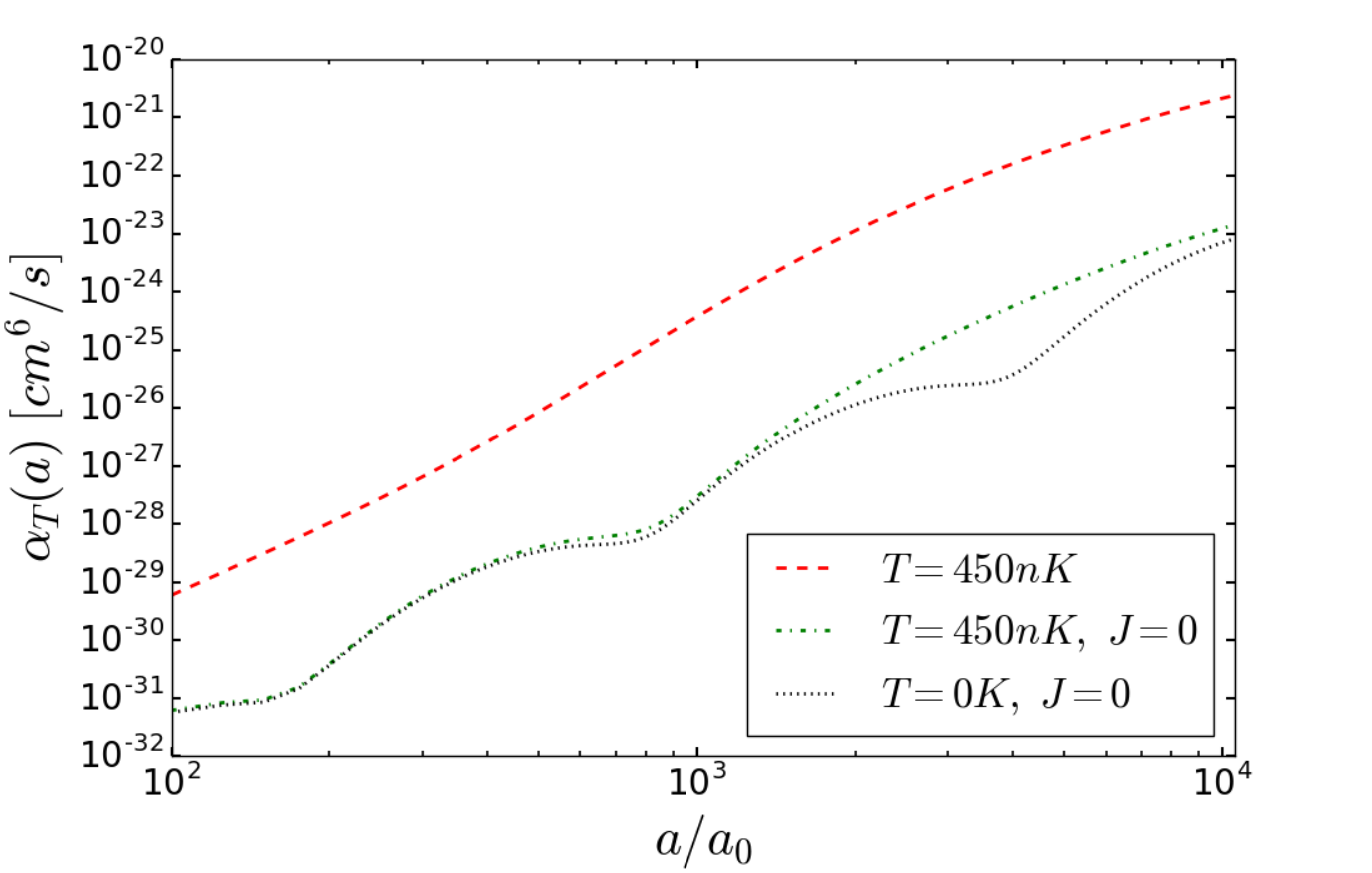}}
	\caption{Shown on the left is the rate constant $\alpha_{T}(a)$ for \KCS,
	with $a_{*0}
		=500a_{0}$ and $\eta_{*}=0.2$, and on the right is the 
		rate constant $\alpha_{T}(a)$ for \LICS, with $a_{*0}
		=805a_{0}$ \cite{Ulmanis2016} and $\eta_{*}=0.2$.}
	\label{fig:alphaExtra}
\end{figure}
A few additional systems have been studied in order to 
facilitate a broader understanding 
of the significance of higher partial wave contributions to the three-body
recombination rate across a range of $A_1$--$A_2$ mass ratios, and the rates for
two of these are presented in Fig. \ref{fig:alphaExtra}. Lacking known values 
for the three-body and inelasticity parameters in the \KCS~system, we set 
$a_{*0}=500a_{0}$ and $\eta_{*}=0.2$. For \LICS, we use 
the three-body parameter $a_{*0}\approx805a_{0}$
obtained from the value $a_{-}=-1777a_0$ \cite{Ulmanis2016} via the universal 
relation between the two parameters. In Ref. \cite{Ulmanis2016}, values 
for $\eta_{*}$ were estimated at 120 and 450 nK and are given by
$\eta_{*}^{(120)}=0.61$ and $\eta_{*}^{(450)}=0.86$, respectively. 
For our purposes, we set $\eta_{*}=0.2$ for this system in order to clearly show
locations of minima and the effects of finite-temperature and higher 
partial-wave contributions. 
The two plots in \fig{alphaExtra} further illustrate the
trend that systems with more extreme mass ratios experience
larger $J\geq1$ recombination rate contributions.

\end{appendix}

\newpage

\begin{thebibliography}{46}
\expandafter\ifx\csname natexlab\endcsname\relax\def\natexlab#1{#1}\fi
\expandafter\ifx\csname bibnamefont\endcsname\relax
  \def\bibnamefont#1{#1}\fi
\expandafter\ifx\csname bibfnamefont\endcsname\relax
  \def\bibfnamefont#1{#1}\fi
\expandafter\ifx\csname citenamefont\endcsname\relax
  \def\citenamefont#1{#1}\fi
\expandafter\ifx\csname url\endcsname\relax
  \def\url#1{\texttt{#1}}\fi
\expandafter\ifx\csname urlprefix\endcsname\relax\def\urlprefix{URL }\fi
\providecommand{\bibinfo}[2]{#2}
\providecommand{\eprint}[2][]{\url{#2}}

\bibitem[{\citenamefont{Hammer and Platter}(2010)}]{Hammer2010a}
\bibinfo{author}{\bibfnamefont{H.-W.} \bibnamefont{Hammer}} \bibnamefont{and}
  \bibinfo{author}{\bibfnamefont{L.}~\bibnamefont{Platter}},
  \bibinfo{journal}{Annu. Rev. Nucl. Part. Sci.} \textbf{\bibinfo{volume}{60}},
  \bibinfo{pages}{207} (\bibinfo{year}{2010}).

\bibitem[{\citenamefont{Efimov}(1970)}]{Efimov70}
\bibinfo{author}{\bibfnamefont{V.}~\bibnamefont{Efimov}},
  \bibinfo{journal}{Phys. Lett.} \textbf{\bibinfo{volume}{33B}},
  \bibinfo{pages}{563} (\bibinfo{year}{1970}).

  
  \bibitem[{\citenamefont{Braaten and
  Hammer}(2006{\natexlab{b}})}]{Braaten2006259}
\bibinfo{author}{\bibfnamefont{E.}~\bibnamefont{Braaten}} \bibnamefont{and}
  \bibinfo{author}{\bibfnamefont{H.-W.} \bibnamefont{Hammer}},
  \bibinfo{journal}{Phys. Rept.} \textbf{\bibinfo{volume}{428}},
  \bibinfo{pages}{259 } (\bibinfo{year}{2006}{\natexlab{b}}).

\bibitem[{\citenamefont{Kraemer et~al.}(2006)\citenamefont{Kraemer, Mark,
  Waldburger, Danzl, Chin, Engeser, Lange, Pilch, Jaakkola, N\"agerl and
  Grimm}}]{Kraemer:2006}
\bibinfo{author}{\bibfnamefont{T.}~\bibnamefont{Kraemer}},
  \bibinfo{author}{\bibfnamefont{M.}~\bibnamefont{Mark}},
  \bibinfo{author}{\bibfnamefont{P.}~\bibnamefont{Waldburger}},
  \bibinfo{author}{\bibfnamefont{J.~G.} \bibnamefont{Danzl}},
  \bibinfo{author}{\bibfnamefont{C.}~\bibnamefont{Chin}},
  \bibinfo{author}{\bibfnamefont{B.}~\bibnamefont{Engeser}},
  \bibinfo{author}{\bibfnamefont{A.~D.} \bibnamefont{Lange}},
  \bibinfo{author}{\bibfnamefont{K.}~\bibnamefont{Pilch}},
  \bibinfo{author}{\bibfnamefont{A.}~\bibnamefont{Jaakkola}},
  \bibinfo{author}{\bibfnamefont{H.-C.} \bibnamefont{N\"agerl}},
  \bibnamefont{and}
  \bibinfo{author}{\bibfnamefont{R.} \bibnamefont{Grimm}}, 
  \bibinfo{journal}{Nature (London)}
  \textbf{\bibinfo{volume}{440}}, \bibinfo{pages}{315} (\bibinfo{year}{2006}).

\bibitem[{\citenamefont{Gross et~al.}(2009)\citenamefont{Gross, Shotan,
  Kokkelmans, and Khaykovich}}]{Gross:2009}
\bibinfo{author}{\bibfnamefont{N.}~\bibnamefont{Gross}},
  \bibinfo{author}{\bibfnamefont{Z.}~\bibnamefont{Shotan}},
  \bibinfo{author}{\bibfnamefont{S.}~\bibnamefont{Kokkelmans}},
  \bibnamefont{and}
  \bibinfo{author}{\bibfnamefont{L.}~\bibnamefont{Khaykovich}},
  \bibinfo{journal}{Phys. Rev. Lett.} \textbf{\bibinfo{volume}{103}},
  \bibinfo{eid}{163202} (\bibinfo{year}{2009}).

\bibitem[{\citenamefont{Pollack et~al.}(2009)\citenamefont{Pollack, Dries, and
  Hulet}}]{pollack:2009}
\bibinfo{author}{\bibfnamefont{S.~E.} \bibnamefont{Pollack}},
  \bibinfo{author}{\bibfnamefont{D.}~\bibnamefont{Dries}}, \bibnamefont{and}
  \bibinfo{author}{\bibfnamefont{R.~G.} \bibnamefont{Hulet}},
  \bibinfo{journal}{Science} \textbf{\bibinfo{volume}{326}},
  \bibinfo{pages}{1683} (\bibinfo{year}{2009}).

\bibitem[{\citenamefont{Gross et~al.}(2010)\citenamefont{Gross, Shotan,
  Kokkelmans, and Khaykovich}}]{Gross2010}
\bibinfo{author}{\bibfnamefont{N.}~\bibnamefont{Gross}},
  \bibinfo{author}{\bibfnamefont{Z.}~\bibnamefont{Shotan}},
  \bibinfo{author}{\bibfnamefont{S.}~\bibnamefont{Kokkelmans}},
  \bibnamefont{and}
  \bibinfo{author}{\bibfnamefont{L.}~\bibnamefont{Khaykovich}},
  \bibinfo{journal}{Phys. Rev. Lett.} \textbf{\bibinfo{volume}{105}},
  \bibinfo{pages}{103203} (\bibinfo{year}{2010}).

\bibitem[{\citenamefont{Ottenstein et~al.}(2008)\citenamefont{Ottenstein,
  Lompe, Kohnen, Wenz, and Jochim}}]{Ottenstein:2008}
\bibinfo{author}{\bibfnamefont{T.~B.} \bibnamefont{Ottenstein}},
  \bibinfo{author}{\bibfnamefont{T.}~\bibnamefont{Lompe}},
  \bibinfo{author}{\bibfnamefont{M.}~\bibnamefont{Kohnen}},
  \bibinfo{author}{\bibfnamefont{A.~N.} \bibnamefont{Wenz}}, \bibnamefont{and}
  \bibinfo{author}{\bibfnamefont{S.}~\bibnamefont{Jochim}},
  \bibinfo{journal}{Phys. Rev. Lett.} \textbf{\bibinfo{volume}{101}},
  \bibinfo{eid}{203202} (\bibinfo{year}{2008}).

\bibitem[{\citenamefont{Huckans et~al.}(2009)\citenamefont{Huckans, Williams,
  Hazlett, Stites, and O'Hara}}]{Huckans:2008fq}
\bibinfo{author}{\bibfnamefont{J.~H.} \bibnamefont{Huckans}},
  \bibinfo{author}{\bibfnamefont{J.~R.} \bibnamefont{Williams}},
  \bibinfo{author}{\bibfnamefont{E.~L.} \bibnamefont{Hazlett}},
  \bibinfo{author}{\bibfnamefont{R.~W.} \bibnamefont{Stites}},
  \bibnamefont{and} \bibinfo{author}{\bibfnamefont{K.~M.}
  \bibnamefont{O'Hara}}, \bibinfo{journal}{Phys. Rev. Lett.}
  \textbf{\bibinfo{volume}{102}}, \bibinfo{pages}{165302}
  (\bibinfo{year}{2009}).

\bibitem[{\citenamefont{Williams et~al.}(2009)\citenamefont{Williams, Hazlett,
  Huckans, Stites, Zhang, and O'Hara}}]{Williams:2009}
\bibinfo{author}{\bibfnamefont{J.~R.} \bibnamefont{Williams}},
  \bibinfo{author}{\bibfnamefont{E.~L.} \bibnamefont{Hazlett}},
  \bibinfo{author}{\bibfnamefont{J.~H.} \bibnamefont{Huckans}},
  \bibinfo{author}{\bibfnamefont{R.~W.} \bibnamefont{Stites}},
  \bibinfo{author}{\bibfnamefont{Y.}~\bibnamefont{Zhang}}, \bibnamefont{and}
  \bibinfo{author}{\bibfnamefont{K.~M.} \bibnamefont{O'Hara}},
  \bibinfo{journal}{Phys. Rev. Lett.} \textbf{\bibinfo{volume}{103}},
  \bibinfo{eid}{130404} (\bibinfo{year}{2009}).
  
\bibitem[{\citenamefont{Floerchinger et~al.}(2009)\citenamefont{Floerchinger,
  Schmidt, and Wetterich}}]{Schmidt:2008fz}
\bibinfo{author}{\bibfnamefont{S.}~\bibnamefont{Floerchinger}},
  \bibinfo{author}{\bibfnamefont{R.}~\bibnamefont{Schmidt}}, \bibnamefont{and}
  \bibinfo{author}{\bibfnamefont{C.}~\bibnamefont{Wetterich}},
  \bibinfo{journal}{Phys. Rev. A} \textbf{\bibinfo{volume}{79}},
  \bibinfo{eid}{053633} (\bibinfo{year}{2009}).

\bibitem[{\citenamefont{Naidon and Ueda}(2009)}]{NU:2009}
\bibinfo{author}{\bibfnamefont{P.}~\bibnamefont{Naidon}} \bibnamefont{and}
  \bibinfo{author}{\bibfnamefont{M.}~\bibnamefont{Ueda}},
  \bibinfo{journal}{Phys. Rev. Lett.} \textbf{\bibinfo{volume}{103}},
  \bibinfo{eid}{073203} (\bibinfo{year}{2009}).

\bibitem[{\citenamefont{Braaten et~al.}(2009)\citenamefont{Braaten, Hammer,
  Kang, and Platter}}]{Braaten:2008wd}
\bibinfo{author}{\bibfnamefont{E.}~\bibnamefont{Braaten}},
  \bibinfo{author}{\bibfnamefont{H.-W.} \bibnamefont{Hammer}},
  \bibinfo{author}{\bibfnamefont{D.}~\bibnamefont{Kang}}, \bibnamefont{and}
  \bibinfo{author}{\bibfnamefont{L.}~\bibnamefont{Platter}},
  \bibinfo{journal}{Phys. Rev. Lett.} \textbf{\bibinfo{volume}{103}},
  \bibinfo{pages}{073202} (\bibinfo{year}{2009}).

\bibitem[{\citenamefont{{Braaten} et~al.}(2010)\citenamefont{{Braaten},
  {Hammer}, {Kang}, and {Platter}}}]{Braaten2010}
\bibinfo{author}{\bibfnamefont{E.}~\bibnamefont{{Braaten}}},
  \bibinfo{author}{\bibfnamefont{H.-W.} \bibnamefont{{Hammer}}},
  \bibinfo{author}{\bibfnamefont{D.}~\bibnamefont{{Kang}}}, \bibnamefont{and}
  \bibinfo{author}{\bibfnamefont{L.}~\bibnamefont{{Platter}}},
  \bibinfo{journal}{\pra} \textbf{\bibinfo{volume}{81}}, \bibinfo{eid}{013605}
  (\bibinfo{year}{2010}).

\bibitem[{\citenamefont{{Hammer} et~al.}(2010)\citenamefont{{Hammer}, {Kang},
  and {Platter}}}]{Hammer2010}
\bibinfo{author}{\bibfnamefont{H.-W.} \bibnamefont{{Hammer}}},
  \bibinfo{author}{\bibfnamefont{D.}~\bibnamefont{{Kang}}}, \bibnamefont{and}
  \bibinfo{author}{\bibfnamefont{L.}~\bibnamefont{{Platter}}},
  \bibinfo{journal}{\pra} \textbf{\bibinfo{volume}{82}}, \bibinfo{eid}{022715}
  (\bibinfo{year}{2010}).

\bibitem[{\citenamefont{Naidon and Endo}(2017)}]{0034-4885-80-5-056001}
\bibinfo{author}{\bibfnamefont{P.}~\bibnamefont{Naidon}} \bibnamefont{and}
  \bibinfo{author}{\bibfnamefont{S.}~\bibnamefont{Endo}},
  \bibinfo{journal}{Rep. Prog. Phys.}
  \textbf{\bibinfo{volume}{80}}, \bibinfo{pages}{056001}
  (\bibinfo{year}{2017}).

\bibitem[{\citenamefont{Barontini et~al.}(2009)\citenamefont{Barontini, Weber,
  Rabatti, Catani, Thalhammer, Inguscio, and Minardi}}]{Barontini:2009}
\bibinfo{author}{\bibfnamefont{G.}~\bibnamefont{Barontini}},
  \bibinfo{author}{\bibfnamefont{C.}~\bibnamefont{Weber}},
  \bibinfo{author}{\bibfnamefont{F.}~\bibnamefont{Rabatti}},
  \bibinfo{author}{\bibfnamefont{J.}~\bibnamefont{Catani}},
  \bibinfo{author}{\bibfnamefont{G.}~\bibnamefont{Thalhammer}},
  \bibinfo{author}{\bibfnamefont{M.}~\bibnamefont{Inguscio}}, \bibnamefont{and}
  \bibinfo{author}{\bibfnamefont{F.}~\bibnamefont{Minardi}},
  \bibinfo{journal}{Phys. Rev. Lett.} \textbf{\bibinfo{volume}{103}},
  \bibinfo{pages}{043201} (\bibinfo{year}{2009}).

\bibitem[{\citenamefont{Tung et~al.}(2014)\citenamefont{Tung,
  Jim\'enez-Garc\'{\i}a, Johansen, Parker, and Chin}}]{PhysRevLett.113.240402}
\bibinfo{author}{\bibfnamefont{S.-K.} \bibnamefont{Tung}},
  \bibinfo{author}{\bibfnamefont{K.}~\bibnamefont{Jim\'enez-Garc\'{\i}a}},
  \bibinfo{author}{\bibfnamefont{J.}~\bibnamefont{Johansen}},
  \bibinfo{author}{\bibfnamefont{C.~V.} \bibnamefont{Parker}},
  \bibnamefont{and} \bibinfo{author}{\bibfnamefont{C.}~\bibnamefont{Chin}},
  \bibinfo{journal}{Phys. Rev. Lett.} \textbf{\bibinfo{volume}{113}},
  \bibinfo{pages}{240402} (\bibinfo{year}{2014}).

\bibitem[{\citenamefont{Pires et~al.}(2014)\citenamefont{Pires, Ulmanis,
  H\"afner, Repp, Arias, Kuhnle, and Weidem\"uller}}]{PhysRevLett.112.250404}
\bibinfo{author}{\bibfnamefont{R.}~\bibnamefont{Pires}},
  \bibinfo{author}{\bibfnamefont{J.}~\bibnamefont{Ulmanis}},
  \bibinfo{author}{\bibfnamefont{S.}~\bibnamefont{H\"afner}},
  \bibinfo{author}{\bibfnamefont{M.}~\bibnamefont{Repp}},
  \bibinfo{author}{\bibfnamefont{A.}~\bibnamefont{Arias}},
  \bibinfo{author}{\bibfnamefont{E.~D.} \bibnamefont{Kuhnle}},
  \bibnamefont{and}
  \bibinfo{author}{\bibfnamefont{M.}~\bibnamefont{Weidem\"uller}},
  \bibinfo{journal}{Phys. Rev. Lett.} \textbf{\bibinfo{volume}{112}},
  \bibinfo{pages}{250404} (\bibinfo{year}{2014}).

\bibitem[{\citenamefont{Helfrich et~al.}(2010)\citenamefont{Helfrich, Hammer,
  and Petrov}}]{Helfrich2010}
\bibinfo{author}{\bibfnamefont{K.}~\bibnamefont{Helfrich}},
  \bibinfo{author}{\bibfnamefont{H.-W.} \bibnamefont{Hammer}},
  \bibnamefont{and} \bibinfo{author}{\bibfnamefont{D.~S.}
  \bibnamefont{Petrov}}, \bibinfo{journal}{Phys. Rev. A}
  \textbf{\bibinfo{volume}{81}}, \bibinfo{pages}{042715}
  (\bibinfo{year}{2010}).

\bibitem[{\citenamefont{Zinner and Nygaard}(2015)}]{Zinner2015}
\bibinfo{author}{\bibfnamefont{N.~T.} \bibnamefont{Zinner}} \bibnamefont{and}
  \bibinfo{author}{\bibfnamefont{N.~G.} \bibnamefont{Nygaard}},
  \bibinfo{journal}{Few-Body Syst.} \textbf{\bibinfo{volume}{56}},
  \bibinfo{pages}{125} (\bibinfo{year}{2015}).

\bibitem[{\citenamefont{Wang et~al.}(2012)\citenamefont{Wang, Wang, D'Incao,
  and Greene}}]{Wang2012}
\bibinfo{author}{\bibfnamefont{Y.}~\bibnamefont{Wang}},
  \bibinfo{author}{\bibfnamefont{J.}~\bibnamefont{Wang}},
  \bibinfo{author}{\bibfnamefont{J.~P.} \bibnamefont{D'Incao}},
  \bibnamefont{and} \bibinfo{author}{\bibfnamefont{C.~H.}
  \bibnamefont{Greene}}, \bibinfo{journal}{Phys. Rev. Lett.}
  \textbf{\bibinfo{volume}{109}}, \bibinfo{pages}{243201}
  (\bibinfo{year}{2012}).

\bibitem[{\citenamefont{Blume and Yan}(2014)}]{PhysRevLett.113.213201}
\bibinfo{author}{\bibfnamefont{D.}~\bibnamefont{Blume}} \bibnamefont{and}
  \bibinfo{author}{\bibfnamefont{Y.}~\bibnamefont{Yan}},
  \bibinfo{journal}{Phys. Rev. Lett.} \textbf{\bibinfo{volume}{113}},
  \bibinfo{pages}{213201} (\bibinfo{year}{2014}).

\bibitem[{\citenamefont{Acharya et~al.}(2016)\citenamefont{Acharya, Ji, and
  Platter}}]{PhysRevA.94.032702}
\bibinfo{author}{\bibfnamefont{B.}~\bibnamefont{Acharya}},
  \bibinfo{author}{\bibfnamefont{C.}~\bibnamefont{Ji}}, \bibnamefont{and}
  \bibinfo{author}{\bibfnamefont{L.}~\bibnamefont{Platter}},
  \bibinfo{journal}{Phys. Rev. A} \textbf{\bibinfo{volume}{94}},
  \bibinfo{pages}{032702} (\bibinfo{year}{2016}).

\bibitem[{\citenamefont{Braaten et~al.}(2008)\citenamefont{Braaten, Hammer,
  Kang, and Platter}}]{Braaten2008}
\bibinfo{author}{\bibfnamefont{E.}~\bibnamefont{Braaten}},
  \bibinfo{author}{\bibfnamefont{H.-W.} \bibnamefont{Hammer}},
  \bibinfo{author}{\bibfnamefont{D.}~\bibnamefont{Kang}}, \bibnamefont{and}
  \bibinfo{author}{\bibfnamefont{L.}~\bibnamefont{Platter}},
  \bibinfo{journal}{Phys. Rev. A} \textbf{\bibinfo{volume}{78}},
  \bibinfo{pages}{043605} (\bibinfo{year}{2008}).

\bibitem[{\citenamefont{Petrov and Werner}(2015)}]{Petrov2015}
\bibinfo{author}{\bibfnamefont{D.~S.} \bibnamefont{Petrov}} \bibnamefont{and}
  \bibinfo{author}{\bibfnamefont{F.}~\bibnamefont{Werner}},
  \bibinfo{journal}{Phys. Rev. A} \textbf{\bibinfo{volume}{92}},
  \bibinfo{pages}{022704} (\bibinfo{year}{2015}).

\bibitem[{\citenamefont{Skorniakov and
  Ter-Martirosian}(1957)}]{Skorniakov:1957aa}
\bibinfo{author}{\bibfnamefont{G.~V.} \bibnamefont{Skorniakov}}
  \bibnamefont{and} \bibinfo{author}{\bibfnamefont{K.~A.}
  \bibnamefont{Ter-Martirosian}}, \bibinfo{journal}{Sov. Phys. JETP}
  \textbf{\bibinfo{volume}{4}}, \bibinfo{pages}{648} (\bibinfo{year}{1957}).

\bibitem[{\citenamefont{Bedaque
  et~al.}(1999{\natexlab{a}})\citenamefont{Bedaque, Hammer, and van
  Kolck}}]{Bedaque:1998km}
\bibinfo{author}{\bibfnamefont{P.~F.} \bibnamefont{Bedaque}},
  \bibinfo{author}{\bibfnamefont{H.-W.} \bibnamefont{Hammer}},
  \bibnamefont{and} \bibinfo{author}{\bibfnamefont{U.}~\bibnamefont{van
  Kolck}}, \bibinfo{journal}{Nucl. Phys. A} \textbf{\bibinfo{volume}{646}},
  \bibinfo{pages}{444} (\bibinfo{year}{1999}{\natexlab{a}}).

\bibitem[{\citenamefont{Ji et~al.}(2012)\citenamefont{Ji, Phillips, and
  Platter}}]{Ji:2011qg}
\bibinfo{author}{\bibfnamefont{C.}~\bibnamefont{Ji}},
  \bibinfo{author}{\bibfnamefont{D.~R.} \bibnamefont{Phillips}},
  \bibnamefont{and} \bibinfo{author}{\bibfnamefont{L.}~\bibnamefont{Platter}},
  \bibinfo{journal}{Ann. Phys. (N.Y.)} \textbf{\bibinfo{volume}{327}},
  \bibinfo{pages}{1803} (\bibinfo{year}{2012}).

\bibitem[{\citenamefont{Helfrich and Hammer}(2011)}]{Helfrich:2011ut}
\bibinfo{author}{\bibfnamefont{K.}~\bibnamefont{Helfrich}} \bibnamefont{and}
  \bibinfo{author}{\bibfnamefont{H.-W.} \bibnamefont{Hammer}},
  \bibinfo{journal}{J. Phys. B} \textbf{\bibinfo{volume}{44}},
  \bibinfo{pages}{215301} (\bibinfo{year}{2011}).

\bibitem[{\citenamefont{Efimov}(1973)}]{Efimov73}
\bibinfo{author}{\bibfnamefont{V.}~\bibnamefont{Efimov}},
  \bibinfo{journal}{Nucl. Phys. A} \textbf{\bibinfo{volume}{210}},
  \bibinfo{pages}{157} (\bibinfo{year}{1973}).

\bibitem[{\citenamefont{Kartavtsev and Malykh}(2008)}]{Kartavtsev2008}
\bibinfo{author}{\bibfnamefont{O.~I.} \bibnamefont{Kartavtsev}}
  \bibnamefont{and} \bibinfo{author}{\bibfnamefont{A.~V.}
  \bibnamefont{Malykh}}, \bibinfo{journal}{JETP Letters}
  \textbf{\bibinfo{volume}{86}}, \bibinfo{pages}{625} (\bibinfo{year}{2008}).

\bibitem[{\citenamefont{Bedaque
  et~al.}(1999{\natexlab{b}})\citenamefont{Bedaque, Hammer, and van
  Kolck}}]{Bedaque:1998kg}
\bibinfo{author}{\bibfnamefont{P.~F.} \bibnamefont{Bedaque}},
  \bibinfo{author}{\bibfnamefont{H.-W.} \bibnamefont{Hammer}},
  \bibnamefont{and} \bibinfo{author}{\bibfnamefont{U.}~\bibnamefont{van
  Kolck}}, \bibinfo{journal}{Phys. Rev. Lett.} \textbf{\bibinfo{volume}{82}},
  \bibinfo{pages}{463} (\bibinfo{year}{1999}{\natexlab{b}}).

\bibitem[{\citenamefont{Hammer and Mehen}(2001)}]{Hammer:2000nf}
\bibinfo{author}{\bibfnamefont{H.-W.} \bibnamefont{Hammer}} \bibnamefont{and}
  \bibinfo{author}{\bibfnamefont{T.}~\bibnamefont{Mehen}},
  \bibinfo{journal}{Nucl. Phys. A} \textbf{\bibinfo{volume}{690}},
  \bibinfo{pages}{535} (\bibinfo{year}{2001}).

\bibitem[{\citenamefont{Hetherington and Schick}(1965)}]{PhysRev.137.B935}
\bibinfo{author}{\bibfnamefont{J.~H.} \bibnamefont{Hetherington}}
  \bibnamefont{and} \bibinfo{author}{\bibfnamefont{L.~H.}
  \bibnamefont{Schick}}, \bibinfo{journal}{Phys. Rev.}
  \textbf{\bibinfo{volume}{137}}, \bibinfo{pages}{B935} (\bibinfo{year}{1965}).

\bibitem[{\citenamefont{Braaten et~al.}(2017)\citenamefont{Braaten, Hammer, and
  Lepage}}]{Braaten:2016dsw}
\bibinfo{author}{\bibfnamefont{E.}~\bibnamefont{Braaten}},
  \bibinfo{author}{\bibfnamefont{H.~W.} \bibnamefont{Hammer}},
  \bibnamefont{and} \bibinfo{author}{\bibfnamefont{G.~P.}
  \bibnamefont{Lepage}}, \bibinfo{journal}{Phys. Rev. A.}
  \textbf{\bibinfo{volume}{95}}, \bibinfo{pages}{012708}
  (\bibinfo{year}{2017}).

\bibitem[{\citenamefont{Braaten et~al.}(2016)\citenamefont{Braaten, Hammer, and
  Lepage}}]{Braaten:2016sja}
\bibinfo{author}{\bibfnamefont{E.}~\bibnamefont{Braaten}},
  \bibinfo{author}{\bibfnamefont{H.~W.} \bibnamefont{Hammer}},
  \bibnamefont{and} \bibinfo{author}{\bibfnamefont{G.~P.}
  \bibnamefont{Lepage}}, \bibinfo{journal}{Phys. Rev. D}
  \textbf{\bibinfo{volume}{94}}, \bibinfo{pages}{056006}
  (\bibinfo{year}{2016}).

\bibitem[{\citenamefont{Esry et~al.}(2001)\citenamefont{Esry, Greene, and
  Suno}}]{Esry2001}
\bibinfo{author}{\bibfnamefont{B.~D.} \bibnamefont{Esry}},
  \bibinfo{author}{\bibfnamefont{C.~H.} \bibnamefont{Greene}},
  \bibnamefont{and} \bibinfo{author}{\bibfnamefont{H.}~\bibnamefont{Suno}},
  \bibinfo{journal}{Phys. Rev. A} \textbf{\bibinfo{volume}{65}},
  \bibinfo{pages}{010705} (\bibinfo{year}{2001}).

\bibitem[{\citenamefont{Bloom et~al.}(2013)\citenamefont{Bloom, Hu, Cumby, and
  Jin}}]{Bloom2013}
\bibinfo{author}{\bibfnamefont{R.~S.} \bibnamefont{Bloom}},
  \bibinfo{author}{\bibfnamefont{M.-G.} \bibnamefont{Hu}},
  \bibinfo{author}{\bibfnamefont{T.~D.} \bibnamefont{Cumby}}, \bibnamefont{and}
  \bibinfo{author}{\bibfnamefont{D.~S.} \bibnamefont{Jin}},
  \bibinfo{journal}{Phys. Rev. Lett.} \textbf{\bibinfo{volume}{111}},
  \bibinfo{pages}{105301} (\bibinfo{year}{2013}).

\bibitem[{\citenamefont{Ulmanis et~al.}(2016)\citenamefont{Ulmanis, H\"afner,
  Pires, Werner, Petrov, Kuhnle, and Weidem\"uller}}]{Ulmanis2016}
\bibinfo{author}{\bibfnamefont{J.}~\bibnamefont{Ulmanis}},
  \bibinfo{author}{\bibfnamefont{S.}~\bibnamefont{H\"afner}},
  \bibinfo{author}{\bibfnamefont{R.}~\bibnamefont{Pires}},
  \bibinfo{author}{\bibfnamefont{F.}~\bibnamefont{Werner}},
  \bibinfo{author}{\bibfnamefont{D.~S.} \bibnamefont{Petrov}},
  \bibinfo{author}{\bibfnamefont{E.~D.} \bibnamefont{Kuhnle}},
  \bibnamefont{and}
  \bibinfo{author}{\bibfnamefont{M.}~\bibnamefont{Weidem\"uller}},
  \bibinfo{journal}{Phys. Rev. A} \textbf{\bibinfo{volume}{93}},
  \bibinfo{pages}{022707} (\bibinfo{year}{2016}).

\bibitem[{\citenamefont{D'Incao and Esry}(2006)}]{DIncao2006}
\bibinfo{author}{\bibfnamefont{J.~P.} \bibnamefont{D'Incao}} \bibnamefont{and}
  \bibinfo{author}{\bibfnamefont{B.~D.} \bibnamefont{Esry}},
  \bibinfo{journal}{Phys. Rev. A} \textbf{\bibinfo{volume}{73}},
  \bibinfo{pages}{030702} (\bibinfo{year}{2006}).

\bibitem[{\citenamefont{Zirbel et~al.}(2008)\citenamefont{Zirbel, Ni,
  Ospelkaus, D'Incao, Wieman, Ye, and Jin}}]{Zirbel2008}
\bibinfo{author}{\bibfnamefont{J.~J.} \bibnamefont{Zirbel}},
  \bibinfo{author}{\bibfnamefont{K.-K.} \bibnamefont{Ni}},
  \bibinfo{author}{\bibfnamefont{S.}~\bibnamefont{Ospelkaus}},
  \bibinfo{author}{\bibfnamefont{J.~P.} \bibnamefont{D'Incao}},
  \bibinfo{author}{\bibfnamefont{C.~E.} \bibnamefont{Wieman}},
  \bibinfo{author}{\bibfnamefont{J.}~\bibnamefont{Ye}}, \bibnamefont{and}
  \bibinfo{author}{\bibfnamefont{D.~S.} \bibnamefont{Jin}},
  \bibinfo{journal}{Phys. Rev. Lett.} \textbf{\bibinfo{volume}{100}},
  \bibinfo{pages}{143201} (\bibinfo{year}{2008}).

\bibitem[{\citenamefont{Wacker et~al.}(2016)\citenamefont{Wacker, J\o{}rgensen,
  Birkmose, Winter, Mikkelsen, Sherson, Zinner, and Arlt}}]{Wacker2016}
\bibinfo{author}{\bibfnamefont{L.~J.} \bibnamefont{Wacker}},
  \bibinfo{author}{\bibfnamefont{N.~B.} \bibnamefont{J\o{}rgensen}},
  \bibinfo{author}{\bibfnamefont{D.}~\bibnamefont{Birkmose}},
  \bibinfo{author}{\bibfnamefont{N.}~\bibnamefont{Winter}},
  \bibinfo{author}{\bibfnamefont{M.}~\bibnamefont{Mikkelsen}},
  \bibinfo{author}{\bibfnamefont{J.}~\bibnamefont{Sherson}},
  \bibinfo{author}{\bibfnamefont{N.}~\bibnamefont{Zinner}}, \bibnamefont{and}
  \bibinfo{author}{\bibfnamefont{J.~J.} \bibnamefont{Arlt}},
  \bibinfo{journal}{Phys. Rev. Lett.} \textbf{\bibinfo{volume}{117}},
  \bibinfo{pages}{163201} (\bibinfo{year}{2016}).

\bibitem[{\citenamefont{Maier et~al.}(2015)\citenamefont{Maier, Eisele,
  Tiemann, and Zimmermann}}]{Maier2015}
\bibinfo{author}{\bibfnamefont{R.~A.~W.} \bibnamefont{Maier}},
  \bibinfo{author}{\bibfnamefont{M.}~\bibnamefont{Eisele}},
  \bibinfo{author}{\bibfnamefont{E.}~\bibnamefont{Tiemann}}, \bibnamefont{and}
  \bibinfo{author}{\bibfnamefont{C.}~\bibnamefont{Zimmermann}},
  \bibinfo{journal}{Phys. Rev. Lett.} \textbf{\bibinfo{volume}{115}},
  \bibinfo{pages}{043201} (\bibinfo{year}{2015}).

\bibitem[{\citenamefont{Ji et~al.}(2015)\citenamefont{Ji, Braaten, Phillips,
  and Platter}}]{Ji:2015hha}
\bibinfo{author}{\bibfnamefont{C.}~\bibnamefont{Ji}},
  \bibinfo{author}{\bibfnamefont{E.}~\bibnamefont{Braaten}},
  \bibinfo{author}{\bibfnamefont{D.~R.} \bibnamefont{Phillips}},
  \bibnamefont{and} \bibinfo{author}{\bibfnamefont{L.}~\bibnamefont{Platter}},
  \bibinfo{journal}{Phys. Rev. A} \textbf{\bibinfo{volume}{92}},
  \bibinfo{pages}{030702} (\bibinfo{year}{2015}).

\end{thebibliography}
\bibliographystyle{apsrev}

\end{document}